\documentclass[12pt]{iopart}
\expandafter\let\csname equation*\endcsname\relax
\expandafter\let\csname endequation*\endcsname\relax
\usepackage{amsmath}
\usepackage{graphicx} 
\usepackage{bm}
\usepackage{siunitx,makecell}
\usepackage{tcolorbox,tabularx,booktabs}
\usepackage{url,hyperref}
\usepackage{multirow}
\newcommand{\Eatom}{E_{\mathrm{atom}}}
\newcommand{\Esolv}{E_{\mathrm{solv}}}
\newcommand{\Etot}{E_{\mathrm{tot}}}
\newcommand{\IE}{\mathrm{IE}}
\newcommand{\EA}{\mathrm{EA}}
\newcommand{\HOMO}{E_{\mathrm{HOMO}}}
\newcommand{\LUMO}{E_{\mathrm{LUMO}}}

\usepackage{pdfpages}

\usepackage{lineno}

\newcommand{\filename}[1]{\mbox{\protect\path{#1}}}
\newcommand{\jsonkeyword}[1]{\protect\path{#1}}

\begin{document}
\title[Calculated solvation and ionization energies]{Calculated solvation and ionization energies for thousands of organic molecules relevant to battery design}
\author{Jan Weinreich\footnote[2]{Ecole Polytechnique F\'{e}d\'{e}rale de Lausanne, Institut des Sciences et Ing\'{e}nierie chimiques, BCH 5312 (B\v{a}t. BCH)
CH-1015 Lausanne}\footnote[1]{Contributed equally to this work}, Konstantin Karandashev\footnote[3]{University of Vienna, Faculty of Physics, Kolingasse 14-16, AT-1090 Wien, Austria}\footnote[4]{\path{konstantin.karandashev@univie.ac.at}}$\dagger$, Daniel Jose Arismendi Arrieta\footnote[5]{{Department of Chemistry-{\AA}ngstr\"om, Uppsala University, Box 538, 75121 Uppsala, Sweden}}, Kersti Hermansson\footnotemark[5], O. Anatole von Lilienfeld\footnote[6]{Vector Institute for Artificial Intelligence, Toronto, ON, M5S 1M1, Canada}\footnote[7]{Departments of Chemistry, Materials Science and Engineering, and Physics, University of Toronto, St. George Campus, Toronto, ON, Canada}\footnote[8]{Machine Learning Group, Technische Universit\"at Berlin and Institute for the Foundations of Learning and Data, 10587 Berlin, Germany}}

\begin{abstract}
We present high-quality reference data for two fundamentally important groups of molecular properties related to a compound's utility as a lithium battery electrolyte. The first property is energy changes associated with charge excitations of molecules, namely ionization potential and electron affinity. They were estimated for 7000 randomly chosen molecules with up to 9 non-hydrogen atoms C, N, O, and F (QM9 dataset) using the DH-HF, DF-HF-CABS, PNO-LMP2-F12, and PNO-LCCSD(T)-F12 methods as implemented in the Molpro software, and the aug-cc-pVTZ basis set. Additionally, we provide the corresponding atomization energies at these levels of theory, as well as the CPU time and disk space used during the calculations. The second property is solvation energies for 39 different solvents, which we estimate for 18361 molecules connected to battery design (Electrolyte Genome Project dataset), 309463 randomly chosen molecules with up to 17 non-hydrogen atoms C, N, O, S, and halogens (GDB17 dataset), as well as 88418 atoms-in-molecules of the ZINC database of commercially available compounds and 37772 atoms-in-molecules  of GDB17. For these calculations we used the COnductor-like Screening MOdel for Real Solvents (COSMO-RS) method; we additionally provide estimates of gas-phase atomization energies, as well as information about conformers considered during the COSMO-RS calculations, namely coordinates, energies, and dipole moments.
\end{abstract}

\maketitle

\section{Introduction}

Identifying candidate molecules for lithium battery electrolyte components is an important part of creating batteries suited for different modes of operation. The discovery of molecules and materials with improved functionality is expected to greatly accelerate with machine learning (ML) algorithms in coming years \cite{Liu_Shi:2020,Jha_Liang:2023,Wang_He:2024}. However, achieving adequate accuracy of new ML models critically depends on the availability of high-quality reference data, both for validation of the new models and of the underlying ML algorithms, and for training new models to guide exploration of chemical space. This motivated us to collect high-quality data for two fundamentally important molecular properties.  The first one is energy changes associated with charge excitations of molecules, namely ionization potential and electron affinity. The second one is solvation energies which are related to how usable a compound is as an additive to lithium battery electrolytes.

Ionization potential (IP) and electron affinity (EA) are related to how readily a molecule enters reduction and oxidation reactions. Although the utility of these quantities in battery design has been put into question \cite{Borodin:2019}, they are often used in preliminary screening for battery electrolyte components \cite{Korth:2014,Cheng_Curtiss:2015,Qu_Persson:2015,Lian_Wu:2019}, where these quantities are evaluated with relatively computationally cheap calculations. Thus, our goal here is to provide accurate reference data to allow the testing of the fidelity of such calculations. When choosing our reference calculation methods, we have avoided dependence on any kind of empirical fits, disqualifying density functional theory methods and limiting our considerations to post-Hartree Fock approaches. We also aimed to balance feasible computational time for medium-sized molecules that could be considered for battery electrolytes (as too large molecules tend to create issues with solvent viscosity and ion conductivity \cite{Borodin_Knap:2015}) with accuracy, drawing our attention to the PNO-LMP2-F12 \cite{Werner_Dornbach:2015,Ma_Werner:2015,Krause_Werner:2019} and PNO-LCCSD(T)-F12 \cite{Schwilk_Werner:2017,Ma_Werner:2017,Ma_Werner:2018,Ma_Werner:2020} methods as implemented in Molpro \cite{MOLPRO-WIREs,MOLPRO-JCP,MOLPRO_brief}. The methods improve upon standard coupled-cluster method \cite{Cizek:1966,Bartlett:1989} with single and double excitations and a perturbative treatment of triple excitation \cite{Raghavachari_Head-Gordon:1989,Bartlett_Noga:1990,Stanton:1997} [CCSD(T)] and second-order M\o{}ller-Plesset perturbation theory \cite{Moller_Plesset:1934,Cremer:2011} (MP2) in several ways \cite{Ma_Werner:2018_review}. For each calculation we also present the corresponding computational time to allow using the data for developing novel multilevel machine learning schemes \cite{Zaspel:2018,Heinen_Lilienfeld:2023}. More calculation details are given in Section~\ref{sec:QM9-IPEA}. The resulting dataset is named ``QM9-IPEA''.

   Calculation of solvation energies are of great importance for our understanding of solute-solvent interactions. In the context of battery research, solvation energies are useful for approximating trends in solubility in a given battery electrolyte \cite{Cheng_Curtiss:2015}, and solubility is a fundamentally important factor regarding how usable a compound is as an electrolyte additive. Our method of choice for estimating these quantities is the COnductor-like Screening MOdel for Real Solvents \cite{cosmo,cosmo2,KLAMT200043,doi:10.1021/ar800187p,doi:10.1146/annurev-chembioeng-073009-100903} (COSMO-RS), which is based on quantum chemical calculations and, when compared to empirical approaches such as Reaction Mechanism Generator group solvation~\cite{Chung_Green:2022}, trades a much higher cost~\cite{Weinreich_Lilienfeld:2022} for smaller number of additional parameters and robustness. The accuracy of COSMO-RS is among the highest for describing equilibrium fluid thermodynamics: for instance, a mean unsigned error of $\SI{0.5}{\kilo\text{cal}\per\mole}$ was achieved for experimental solvation energies \cite{doi:10.1021/acs.jcim.6b00081} making it popular for chemical engineering applications \cite{doi:10.1021/ie049139z}. In the context of battery materials research, COSMO-RS also demonstrated reasonable mean absolute errors (as low as 0.14~V) when predicting experimental values of redox potentials \cite{doi:10.1021/acs.jctc.2c00919}.

The balance of accuracy and computational cost provided by COSMO-RS allowed us to construct a dataset with a significant chemical variety of both solvated molecules and the solvents they are solvated in, while accurately reflecting physical trends over such a wide chemical space. We cover a broad spectrum of organic chemistry with our database, with a particular focus on electrolyte applications. The foundation of our database is the Electrolyte Genome Project (EGP) dataset \cite{Qu_Persson:2015}, which we have expanded by randomly selecting structures from the GDB17 \cite{Ruddigkeit_Reymond:2012} database, as well as with fragment structures of AGZ7 \cite{agz7} designed to cover the chemical space of ZINC and GDB17. Details are given in Section~\ref{sec:solquest_details}. Our resulting dataset contains data on 418185 molecules and is named ``SolQuest''. 

\section{QM9-IPEA dataset details}
\label{sec:QM9-IPEA}

\subsection{Molecules included}

The calculations were performed for a subset of the QM9 \cite{Ruddigkeit_Reymond:2012,Ramakrishnan_Lilienfeld:2014} dataset, which is often used for benchmarking in ML algorithm studies. We randomly chose 7000 molecules from the dataset and performed calculations for their states with charges 0, 1, and -1 using the geometries present in the dataset. We used aug-cc-pVTZ \cite{Dunning:1989,Kendall_Harrison:1992,Woon_Dunning:1993} basis set. When running Molpro we set the following thresholds: $10^{-8}$ a.u. for energy convergence, $10^{-7}$ for orthonormality check, $5\cdot 10^{-10}$ for smallest allowed eigenvalue of the overlap matrix.

Note that calculations for 4000 of the QM9 molecules considered in this work previously appeared in Ref.~\cite{Heinen_Lilienfeld:2023} along with 4000 calculations for EGP \cite{Qu_Persson:2015}. We had generated the latter with exactly the same methodology, except we additionally used def2-TZVPP \cite{Weigend_Ahlrichs:2005} basis set for Li, Be, and Ca atoms and aug-cc-pVTZ-PP \cite{Peterson_Dolg:2003,Figgen_Stoll:2005,Peterson_Puzzarini:2005} basis set for Zn and Br atoms.

\subsection{IP and EA calculations}

As mentioned in the Introduction, the local correlation treatment in the PNO-LMP2-F23 and PNO-LCCSD(T)-F12 methods used here improve on the CCSD(T) method. Firstly, the scaling of the methods' cost with system size is improved by using their localized versions [LMP2 \cite{Pulay_Saebo:1986} and LCCSD(T) \cite{Hampel_Werner:1996,Schutz_Werner:2000}] that only consider excitations between localized orbitals that are positioned close to each other, with such close orbitals chosen via pair natural orbital \cite{Ahlrichs_Driessler:1975,Taylor:1981,Staemmler_Jaquet:1981} (PNO) formalism and calculations additionally simplified in a way conceptually  similar to ``domain based local PNO-CCSD'' of Refs.~\cite{Riplinger_Neese:2013a} and~\cite{Riplinger_Neese:2013b}. Secondly, since Slater determinant expansions are not well suited to reproduce cusps at points where the distance between two electrons is zero, the electron wavefunctions include a Slater-type correlation factor with the F12 approach \cite{Ten-no:2004a,Ten-no:2004b}. For the LCCSD(T) methods we used the F12b approximation \cite{Adler_Werner:2007,Knizia_Werner:2009} of the F12 correction as it is the one recommended in Molpro guidelines for larger basis sets. Apart from PNO-LCCSD(T)-F12b energies we also calculated PNO-LCCSD(T*)-F12b energies, where ``(T*)'' indicates using an F12-specific rescaling of the perturbative triples correction described in Ref.~\cite{Knizia_Werner:2009}. Each energy evaluation with these methods leaves as byproducts energies obtained with Hartree-Fock (HF) and HF with complementary auxiliary basis sets singles correction \cite{Adler_Werner:2007,Knizia_Werner:2008} (HF-CABS). Correcting these values for the wrong cusp behavior with a density function (DF) based model \cite{Giner_Toulouse:2018,Loos_Giner:2019,Giner_Toulouse:2020} yields DH-HF and DH-HF-CABS results presented in this work. For calculating open shell species (charged molecules and individual atoms) we used restricted versions of DH-HF, DH-HF-CABS, and PNO-LMP2-F12 [DF-RHF, DFRHF-CABS, and PNO-LRMP2-F12 \cite{Krause_Werner:2019}] and unrestricted versions of all Coupled Cluster methods \cite{Ma_Werner:2020} [PNO-UCCSD-F12b, PNO-UCCSD(T)-F12b, and PNO-UCCSD(T*)-F12b].

Lastly, since there are two differing definitions for EA found in the literature, we note that in this manuscript we define IP (or ionization energy IE)  and EA as energies required to detach one electron from a molecule $\mathrm{X}$ when its charge is 0 or -1
\begin{align}
\mathrm{X} + \IE &\rightarrow \mathrm{X}^{+}+e^{-}\\
\mathrm{X}^{-} + \EA &\rightarrow \mathrm{X}+e^{-},
\end{align}
which are the definitions consistent with Ref.~\cite{McNaught_Wilkinson:1997}. In other words, the quantities are defined as
\begin{align}
\IE&:=E(1)-E(0),\label{eq:IE_definition}\\
\EA&:=E(0)-E(-1),\label{eq:EA_definition}
\end{align}
where $E(-1)$, $E(0)$, and $E(1)$ are total energies of the molecule at charges $-1$, $0$, and $1$. 

\section{SolQuest dataset details}
\label{sec:solquest_details}

\subsection{Molecules included}

As mentioned in the introduction, we performed calculations for compounds represented by Simplified Molecular Input Line Entry System \cite{Weininger:1988} (SMILES) and collected from several sources, namely the entire EGP dataset \cite{Qu_Persson:2015} of molecules connected to battery design, randomly selected compounds from the GDB17 \cite{Ruddigkeit_Reymond:2012} dataset of molecules containing up to 17 non-hydrogen atoms (C, N, O, S, and halogens), as well as AGZ7 \cite{agz7}.  The latter is the complete set of {\bf a}tom-in-molecule-based fragments \cite{Huang_Lilienfeld:2020} (from now on refered to as \emph{amons}) for {\bf G}DB17 and {\bf Z}INC databases restricted to no more than {\bf 7} non-hydrogen atoms; the resulting molecules include elements H, B, C, N, O, F, Si, P, S, Cl, Br, Sn, and I.

\subsection{Solvation free energy calculations}

To keep the computational costs of creating such an extensive dataset feasible, we turned our attention to continuum solvation models \cite{Tomasi_Cammi:2005}, in particular the COnductor-like Screening MOdel \cite{cosmo} (COSMO) family of solvation methods, which combined with the COSMOtherm \cite{therm} workflow only require a molecule's SMILES for the calculation. COSMO-RS creates a solvation cavity around the solute and models solvent polarization using surface charges. These charges are derived from the solute's electron density, which is obtained from \textit{ab initio} calculations in the solvent. Additionally, COSMO-RS provides insights into hydrogen bonding through charge distribution, treating the solvent as a uniform dielectric medium. The total solvation free energy  is evaluated based on all interactions between surface segments of solvated molecules, incorporating the likelihood of their contact, but instead of sampling individual molecule arrangements, it uses thermodynamic averages for the segments, leading to a self-consistent equation for the chemical potential \cite{cosmo2}. By incorporating corrections for more realistic solvation simulations, COSMO-RS can model the effects of hydrogen bonds \cite{KLAMT200043}. These corrections include fictitious van der Waals interactions, which are proportional to the solute's cavity surface area, addressing the main limitations of the solvent continuum assumption. 

In the COSMOtherm workflow we utilized results from density functional theory calculations conducted with Turbomole \cite{TURBOMOLE}. These calculations employed the B-P86 functional \cite{bp86,assesment} and the def2-TZVPD basis set \cite{metz2000a,Peterson_Dolg:2003, Weigend_Ahlrichs:2005,rappoport2010a}. To ensure a comprehensive dataset, we used COSMOconf \cite{COSMOconf_software} for conformer generation. This tool features predefined procedures specifically designed to produce the most relevant conformers for COSMO-RS applications, beginning with force field-based generation, followed by clustering and diversity-based selection. 

We note that COSMO-RS calculations are based on considering the most important conformers (\emph{i.e.} local minima of potential energy), avoiding extensive sampling associated with  approaches based on Monte Carlo or molecular dynamics simulations \cite{Hansen_Gunsteren:2014}, though also limiting themselves to approximate representation  of a molecule's Boltzmann ensemble. It also means that statistical error of the method is negligible, though theoretically present, as COSMOconf is based on the Balloon algorithm \cite{Vainio_Johnson:2007}, which is a genetic algorithm \cite{Holland:1975}, making it theoretically possible for some important conformers to be missing from COSMOconf's output due to random factors. Lastly, since Balloon generates conformers while accounting for predefined configurations of molecular stereocenters and all molecular SMILES constituting SolQuest defined a single stable enantiomer, all SolQuest calculation results correspond to the one enantiomer defined by the SMILES.

\section{Data overview}

The data is uploaded as several JavaScript Object Notation (JSON) files whose structure is discussed below. For QM9-IPEA, we additionally uploaded a compressed folder with raw Molpro input and output files in case dataset users decide to extract intermediate quantities not considered in this work. We also note that in all uploaded files we used SMILES as a molecule's identifier; while alternative string representations could be more useful from a machine learning perspective \cite{Boyle_Dalke:2018,Krenn_Aspuru-Guzik:2020}, we left their generation up to the potential end user.

\subsection{QM9-IPEA}
\label{subsec:overview_qm9ipea}

The data are kept in two JSON files, \filename{QM9IPEA.json} and \filename{QM9IPEA_atom_ens.json}. The former summarizes all Molpro calculations run for QM9 geometries, the latter provides atom energies necessary to recover atomization energies $\Eatom$; the meaning of different keywords in these files is summarized in Table~\ref{tab:qm9ipea_keywords}. We chose to include ionization energies $\IE$s instead of IPs (which can be trivially recovered from $\IE$s) to keep all energy-related quantities in consistent units (Hartrees). \jsonkeyword{CPU_time} entries contain steps corresponding to individual method calculations, as well as steps corresponding to program operation: \jsonkeyword{INT} (calculating integrals over basis functions relevant for the calculation), \jsonkeyword{FILE} (dumping intermediate data to restart file), and \jsonkeyword{RESTART} (importing restart data). The latter two steps appeared since we reused relevant integrals calculated for neutral species in charged species' calculations; we also used restart functionality to use HF density matrix obtained for the neutral species as the initial density matrix guess for the HF calculation for charged species. Not a number \jsonkeyword{NaN} value of a quantity means that the corresponding calculation or calculation step failed to complete. Note that the CPU times were measured while parallelizing on 12 cores and were not adjusted to single-core; they were observed on AMD Epyc 7,402 processors (24 cores, 512GB of RAM).

\begin{table*}
\begin{tabular}{ll}
\toprule
keyword & description
\\
\hline
\multicolumn{2}{c}{\filename{QM9IPEA.json}}\\
\hline
\jsonkeyword{COORDS} & atom coordinates in Angstroms \\
\jsonkeyword{SYMBOLS} & atom element symbols \\
\jsonkeyword{ENERGY} & total energies for each charge (0, -1, 1) and method considered \\
\jsonkeyword{CPU_TIME} & \multirow{2}{*}{\begin{tabular}{l}CPU times (in seconds) spent at each step of each part \\ of the calculation\end{tabular}} \\
\\
\jsonkeyword{DISK_USAGE} & highest total disk usage in GB \\
\jsonkeyword{ATOMIZATION_ENERGY} & atomization energy at charge 0 (all methods)\\
\jsonkeyword{IONIZATION_ENERGY} & ionization energy for all methods\\
\jsonkeyword{ELECTRON_AFFINITY} & electron affinity for all methods\\
\jsonkeyword{HOMO_ENERGY} & HOMO energy from DFHF calculations\\ 
\jsonkeyword{LUMO_ENERGY} & LUMO energy from DFHF calculations\\ 
\jsonkeyword{QM9_ID} & ID of the molecule in the QM9 dataset \\
\hline
\multicolumn{2}{c}{\filename{QM9IPEA_atom_ens.json}}\\
\hline
\jsonkeyword{SPINS} & the spin assigned to elements during calculations of atomic energies \\
\jsonkeyword{ENERGY} & energies of atoms using different methods\\

\bottomrule
\end{tabular}
\caption{Meaning of quantities found at keywords in \filename{QM9IPEA.json} and \filename{QM9IPEA_atom_ens.json} files. All energies are given in Hartrees with not a number (\jsonkeyword{NaN}) indicating the calculation failed to converge.}
\label{tab:qm9ipea_keywords}
\end{table*}

\begin{figure*}
\center
\includegraphics[width=1.\textwidth]{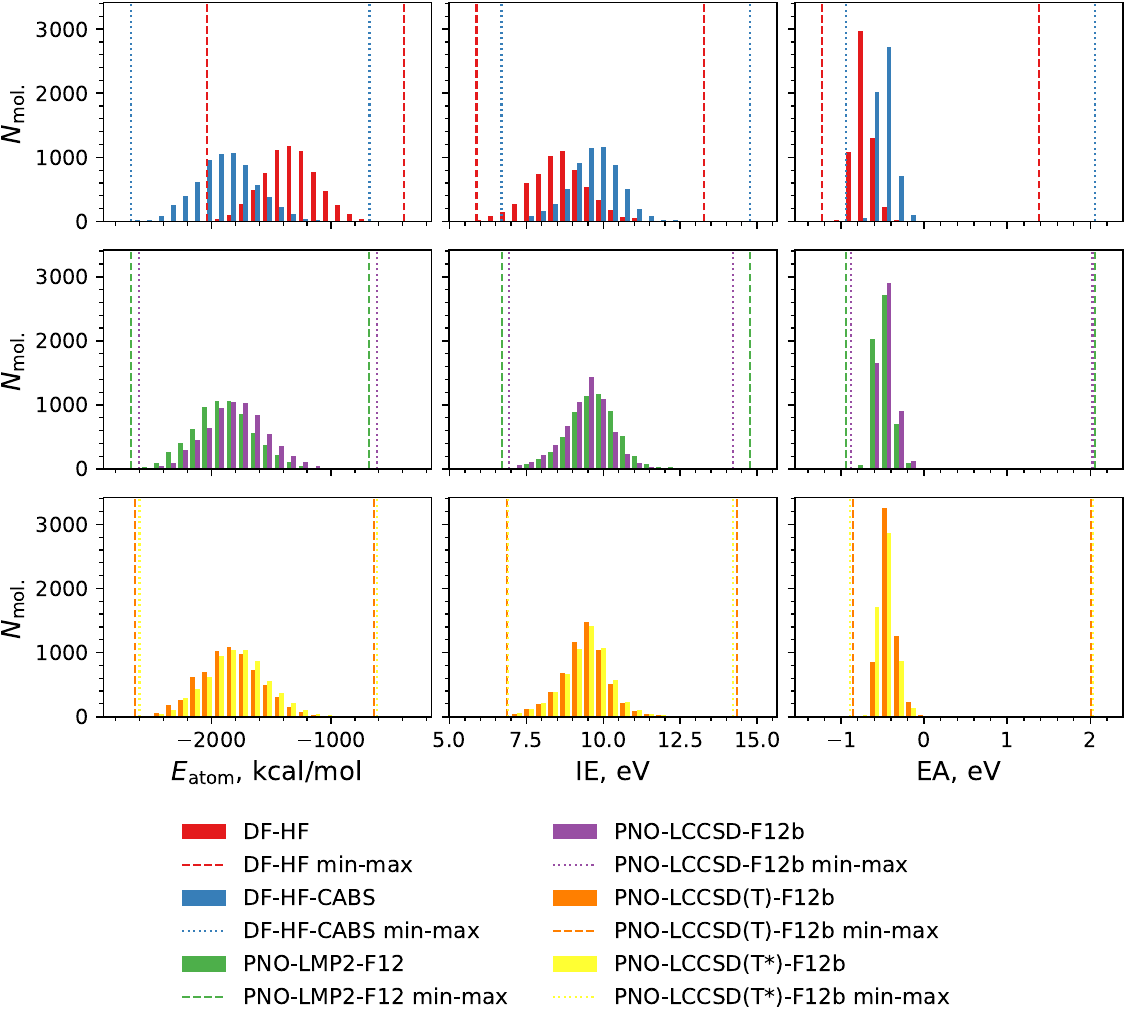}
          \caption{QM9-IPEA molecules' distributions of atomization energy $\Eatom$, ionization energy $\IE$, and electron affinity $\EA$, as evaluated by different methods considered in this work, along with their lowest and highest values ("min-max"). The latter are listed in Supplementary Data.}
     \label{fig:QM9IPEA_distributions} 
\end{figure*}

Distributions of main quantities of interest listed in QM9-IPEA, namely $\Eatom$, $\IE$, and $\EA$, are presented in Figure~\ref{fig:QM9IPEA_distributions}. We observe a significant difference between distributions observed for DF-HF and the other methods, whose distributions in turn look relatively similar. As detailed in Supplementary Data, fitting estimates of these quantities obtained with one method as a linear function of another method yields high $R^{2}$ scores, which are larger than $0.99$ when such a comparison is done between PNO-LCCSD-F12b and PNO-LCCSD(T*)-F12b. This means that, for example, comparing which of two molecules has a lower or higher $\IE$ or $\EA$ can be done with relative certainty at PNO-LCCSD-F12b level of theory already, with the triple excitation contributions largely canceling out.

\begin{figure*}
\center
\includegraphics[width=1.\textwidth]{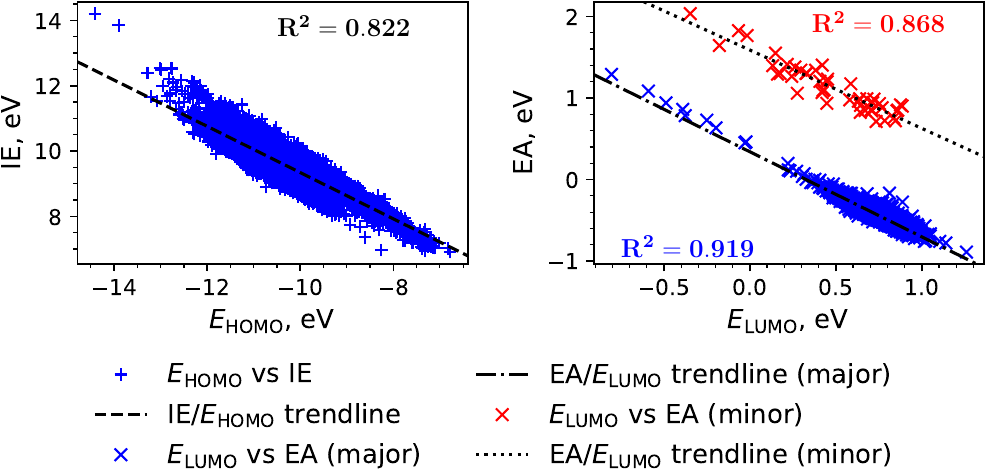}
          \caption{Correlations between ionization energy $\IE$ and HOMO energy $\HOMO$ and between electron affinity $\EA$ and LUMO energy $\LUMO$; $\IE$ and $\EA$ are calculated with PNO-LCCSD(T*)-F12b. Also shown are the trendlines (see Supplementary Data for their expressions) and their $ R^{2}$ factors. As discussed in Subsec.~\ref{subsec:overview_qm9ipea}, the $\EA/\LUMO$ correlation plot breaks QM9-IPEA into two subsets for which separate trendlines are displayed.}
     \label{fig:QM9IPEA_Koopman_brief} 
\end{figure*}

Lastly, we checked how well $\IE$ and $\EA$ correlate with HOMO and LUMO energies ($\HOMO$ and $\LUMO$) obtained at the DF-HF level, a connection implied by Koopman's theorem \cite{Jensen:2007}. The resulting plots look similar between all \emph{ab initio} methods considered in this work and are presented in Supplementary Data, with the plot for PNO-LCCSD(T*)-F12B displayed in Figure~\ref{fig:QM9IPEA_Koopman_brief} as an example. $\IE$ can be fitted well with a linear function of $\HOMO$, while plotting $\EA $ against $\LUMO$ values divides QM9-IPEA into two subsets (the larger dubbed ``major'' and the smaller dubbed ``minor''), each being a good fit for a separate linear trendline. The observation is documented thoroughly in Supplementary Data; finding a reason for it was beyond the scope of this research, although we note in passing its superficial similarity to how in Ref.~\cite{Montavon_Lilienfeld:2013} plotting $\EA$ vs. $\LUMO$ yielded pronounced clustering of molecules while plotting $\IE$ vs. $\HOMO$ did not. We also note that the majority of calculated $\EA$ values (close to $99\%$ for all methods) are negative, implying the basis was insufficiently large to evaluate them accurately;\footnote{In the limit of a complete basis set, if bringing an electron close to a molecule requires energy the former will relax into a state infinitely far away from the latter. Therefore with a complete basis set the calculated $\EA$ is never negative.} however, we hope even negative values could be useful for qualitative molecular ranking or benchmarking. 

\subsection{SolQuest}
\label{subsec:overview_SolQuest}

The dataset is presented in four JSON files listed in Table~\ref{tab:dataset_files}; they can be divided into files for full molecules of EGP and GDB17 and files for amons of GDB17 and ZINC. They are structured differently as amon entries are sorted by
the number of heavy atoms in the amon (\emph{e.g.}, all amons with 3 heavy atoms are stored in \jsonkeyword{ni3}). Because of the large number of amons with 6 or 7 heavy atoms, they are further split into \jsonkeyword{ni6_1}, \jsonkeyword{ni6_2}, etc. Apart from the calculation data and SMILES represeting the molecules the JSON files also contain Extended Connectivity Fingerprints \cite{Rogers_Hahn:2010} with 4 as bond radius (ECFP4) representation vectors to make them more readily usable for machine learning applications. The data is stored behind keywords listed in Table~\ref{tab:solvation_keywords}. For each compound, solvation energies behind the \jsonkeyword{SOLVATION} keyword additionally have one of the solvent keywords listed in Supplementary Data. The \jsonkeyword{ENERGY} keyword denotes Boltzmann average of energy without solvent over configurations used in the solvation energy calculations. Note that the number of entries in \filename{EGP.json}, \filename{AMONS_GDB17.json}, and \filename{AMONS_ZINC.json} is smaller than the number of molecules in the datasets from which they were taken because we excluded molecules for which the calculations failed. The difference between number of entries and number of molecules is due to repetitions of the same molecule inside and between the four subsets of SolQuest; the numbers of unique molecules were obtained by comparing canonical SMILES generated by RdKit \cite{software:RDKit}.

\begin{table*}
\begin{tabular}{llcc}
\toprule
file name & description & num. entries & num. molecules \\
\midrule
\filename{AMONS_GDB17.json} & GDB17 amons & \phantom{0}37860 & \phantom{0}37772\\
\filename{AMONS_ZINC.json} & ZINC amons & \phantom{0}88771 & \phantom{0}88418 \\
\filename{GDB17.json} & subset of GDB17 & 309468 & 309463 \\
\filename{EGP.json} & EGP molecules & \phantom{0}18362 & \phantom{0}18361\\
\midrule
\_ & total & 454461 & 453450\\
\bottomrule
\end{tabular}
\caption{Names of files containing SolQuest data, along with number of entries and non-repeating molecules that each file contains, and the total number of entries and non-repeating molecules.}
\label{tab:dataset_files}
\end{table*}

\begin{table*}
\begin{tabular}{ll}
\toprule
keyword & description \\
\hline
\jsonkeyword{ECFP} & ECFP4 representation vector \\
\jsonkeyword{SMILES} & SMILES string \\
\jsonkeyword{SYMBOLS} & atomic symbols \\
\jsonkeyword{COORDS} & atomic positions for each conformer in Angstrom \\
\jsonkeyword{ATOMIZATION} & atomization energy of each conformer in kcal/mol \\
\jsonkeyword{DIPOLE} & dipole moments and dipole vectors, both for each conformer, in Debye \\
\jsonkeyword{ENERGY} & average energy in Hartree\\
\jsonkeyword{SOLVATION} & solvation energies in kcal/mol for different solvents at 300 K\\
\bottomrule
\end{tabular}
\caption{Meaning of quantities found at keywords in JSON files with the COSMO-RS results.}
\label{tab:solvation_keywords}
\end{table*}

\begin{figure*}
\center
\includegraphics[width=1.\textwidth]{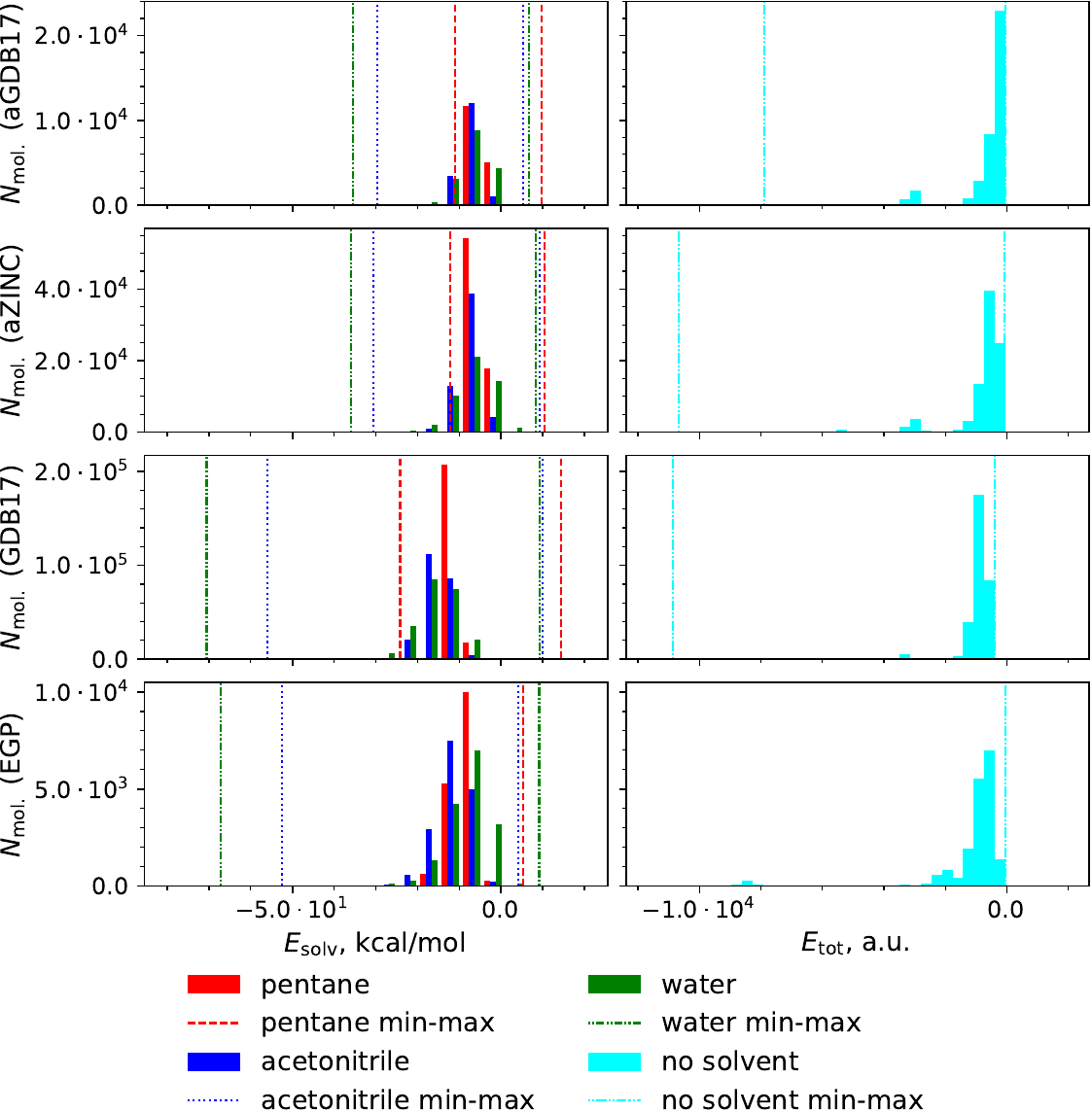}
          \caption{SolQuest molecules' distributions of solvation energies $\Esolv$ for pentane, acetonitrile, and water, as well as total energies without solvent $\Etot$ (see explanation of \jsonkeyword{ENERGY} keyword in Subsec.~\ref{subsec:overview_SolQuest} for definition); vertical lines denote lowest and highest ("min-max") observed values for each quantity, which are listed in Supplementary Data. The distributions are presented for the four subsets of SolQuest based on the dataset the SMILES had been taken from; for brevity "aGDB17" and "aZINC" denote "GDB17 amons" and "ZINC amons".}
     \label{fig:SolQuest_distributions} 
\end{figure*}

We plot distributions of solvation energies $\Esolv$ in water, pentane, and acetonitrile (chosen as the most and the least polar solvents, and the solvent whose dielectric constant is closest to the middle between the ones of water and pentane) along with total energy $\Etot$ (values behind the \jsonkeyword{ENERGY} keyword) in Figure~\ref{fig:SolQuest_distributions}. As expected, solvation energy values tend to become more spread out as the polarity of the solvent increases. We also see that among the four subsets of SolQuest's molecules EGP molecules were the most diverse in terms of distribution of both $\Esolv$ and $\Etot$. The minimum and maximum calculated solvation energies (full information about them presented in Supplementary Data) indicate presence of outliers for which COSMO-RS calculations seemingly broke down, yielding unreasonable solvation energies; we decided to not exclude such points from the dataset and leave it up to the end user to decide whether they are useful.

\section{Conclusions}

In this work we present two comprehensive datasets of interest to battery materials community: QM9-IPEA, which is focused on accurate ionization potentials and electron affinities, and SolQuest, which contains accurate solvation energies for a large number of solvents and a diverse set of molecules. Both datasets include additional information such as CPU time and disk usage (for QM9-IPEA) and details of conformers considered (coordinates, energies, dipole moments) for the solvation energies (for SolQuest). QM9-IPEA provides values of several quantities of interest generated at different levels of theory, potentially helping not only to test multilevel ML approaches, but also to observe which level of theory would be sufficient for such tasks as ranking molecules by their $\IE$ or $\EA$ values (where a degree of error cancellation can occur, as noted in particular for comparing PNO-LCCSD-F12b and PNO-LCCSD(T*)-F12b results). While we raise some concerns about $\EA$ values calculated in this work we hope they can still provide insights about how molecules and electrons interact.

All in all, the data provides a valuable resource for machine learning applications, offering a robust testing ground for novel ML approaches in materials design.

\section{Supplementary Data}

The manuscript's Supplementary Data contains information about quantity extrema referred to in Figures~\ref{fig:QM9IPEA_distributions} and~\ref{fig:SolQuest_distributions}, details of how $\Eatom$, $\IE$, $\EA$ estimates with different methods correlate between each other, detailed information about correlation plots between $\IE$/$\EA$ and $\HOMO$/$\LUMO$, and a full list of solvent keywords for SolQuest's solvation energies.

\section{Data availability}

The GitHub repository containing all scripts that were used to generate and process the data can be found at \url{https://github.com/chemspacelab/VienUppDa}, with the processed data uploaded to Zenodo \cite{zenodo_upload:2024} at \url{https://zenodo.org/records/15252439}.

\ack
This project has received funding from the European Union’s Horizon 2020 research and innovation programme under grant agreement No~957189 (BIG-MAP) and  No~957213 (BATTERY 2030+). O.A.v.L. has received funding from the European Research Council (ERC) under the European Union’s Horizon 2020 research and innovation programme (grant agreement No.~772834). O.A.v.L. has
received support as the Ed Clark Chair of Advanced Materials and as a Canada CIFAR AI Chair. O.A.v.L. acknowledges that this research is part of the University of Toronto’s Acceleration Consortium, which receives funding from the Canada First Research Excellence Fund (CFREF). Obtaining the presented computational results has been facilitated using the queueing system implemented at \href{https://leruli.com}{http://leruli.com}. The project has been supported by the Swedish Research Council (Vetenskapsrådet), and the Swedish National Strategic e-Science program eSSENCE as well as by computing resources from the Swedish National Infrastructure for Computing (SNIC/NAISS). The computational results presented have been achieved using the Vienna Scientific Cluster (VSC).

\section*{References}
\bibliographystyle{iopart-num}
\bibliography{references}

\providecommand{\newblock}{}
\begin{thebibliography}{10}
\expandafter\ifx\csname url\endcsname\relax
  \def\url#1{{\tt #1}}\fi
\expandafter\ifx\csname urlprefix\endcsname\relax\def\urlprefix{URL }\fi
\providecommand{\eprint}[2][]{\url{#2}}

\bibitem{Liu_Shi:2020}
Liu Y, Guo B, Zou X, Li Y and Shi S 2020 {\em Energy Storage Mater.\/} {\bf 31}
  434--450 ISSN 2405-8297
  \urlprefix\url{https://doi.org/10.1016/j.ensm.2020.06.033}

\bibitem{Jha_Liang:2023}
Jha S, Yen M, Soto Y~S, Palmer E, Villafuerte J and Liang H 2023 {\em J. Mater.
  Chem. A\/} {\bf 11}(8) 3904--3936
  \urlprefix\url{http://dx.doi.org/10.1039/D2TA07148G}

\bibitem{Wang_He:2024}
Wang Z, Wang L, Zhang H, Xu H and He X 2024 {\em Nano Converg.\/} {\bf 11}(1) 8
  \urlprefix\url{https://doi.org/10.1186/s40580-024-00417-6}

\bibitem{Borodin:2019}
Borodin O 2019 {\em Curr. Opin. Electrochem.\/} {\bf 13} 86--93
  \urlprefix\url{https://doi.org/10.1016/j.coelec.2018.10.015}

\bibitem{Korth:2014}
Korth M 2014 {\em Phys. Chem. Chem. Phys.\/} {\bf 16} 7919--7926
  \urlprefix\url{https://doi.org/10.1039/C4CP00547C}

\bibitem{Cheng_Curtiss:2015}
Cheng L, Assary R~S, Qu X, Jain A, Ong S~P, Rajput N~N, Persson K and Curtiss
  L~A 2015 {\em J. Phys. Chem. Lett.\/} {\bf 6} 283–291
  \urlprefix\url{https://doi.org/10.1021/jz502319n}

\bibitem{Qu_Persson:2015}
Qu X, Jain A, Rajput N~N, Cheng L, Zhang Y, Ong S~P, Brafman M, Maginn E,
  Curtiss L~A and Persson K~A 2015 {\em Comput. Mater. Sci.\/} {\bf 103} 56--67
  \urlprefix\url{https://doi.org/10.1016/j.commatsci.2015.02.050}

\bibitem{Lian_Wu:2019}
Lian C, Liu H, Li C and Wu J 2019 {\em AIChE J.\/} {\bf 65} 804--810
  \urlprefix\url{https://doi.org/10.1002/aic.16467}

\bibitem{Borodin_Knap:2015}
Borodin O, Olguin M, Spear C~E, Leiter K~W and Knap J 2015 {\em
  Nanotechnology\/} {\bf 26} 354003
  \urlprefix\url{https://doi.org/10.1088/0957-4484/26/35/354003}

\bibitem{Werner_Dornbach:2015}
Werner H~J, Knizia G, Krause C, Schwilk M and Dornbach M 2015 {\em J. Chem.
  Theory Comput.\/} {\bf 11} 484--507 pMID: 26580908
  \urlprefix\url{https://doi.org/10.1021/ct500725e}

\bibitem{Ma_Werner:2015}
Ma Q and Werner H~J 2015 {\em J. Chem. Theory Comput.\/} {\bf 11} 5291--5304
  pMID: 26574323 (\textit{Preprint}
  \eprint{https://doi.org/10.1021/acs.jctc.5b00843})
  \urlprefix\url{https://doi.org/10.1021/acs.jctc.5b00843}

\bibitem{Krause_Werner:2019}
Krause C and Werner H~J 2019 {\em J. Chem. Theory Comput.\/} {\bf 15} 987--1005
  (\textit{Preprint} \eprint{https://doi.org/10.1021/acs.jctc.8b01012})
  \urlprefix\url{https://doi.org/10.1021/acs.jctc.8b01012}

\bibitem{Schwilk_Werner:2017}
Schwilk M, Ma Q, K\"{o}ppl C and Werner H~J 2017 {\em J. Chem. Theory
  Comput.\/} {\bf 13} 3650--3675 pMID: 28661673
  \urlprefix\url{https://doi.org/10.1021/acs.jctc.7b00554}

\bibitem{Ma_Werner:2017}
Ma Q, Schwilk M, K\"{o}ppl C and Werner H~J 2017 {\em J. Chem. Theory
  Comput.\/} {\bf 13} 4871--4896 pMID: 28898081
  \urlprefix\url{https://doi.org/10.1021/acs.jctc.7b00799}

\bibitem{Ma_Werner:2018}
Ma Q and Werner H~J 2018 {\em J. Chem. Theory Comput.\/} {\bf 14} 198--215
  pMID: 29211961 \urlprefix\url{https://doi.org/10.1021/acs.jctc.7b01141}

\bibitem{Ma_Werner:2020}
Ma Q and Werner H~J 2020 {\em J. Chem. Theory Comput.\/} {\bf 16} 3135--3151
  pMID: 32275428 \urlprefix\url{https://doi.org/10.1021/acs.jctc.0c00192}

\bibitem{MOLPRO-WIREs}
Werner H~J, Knowles P~J, Knizia G, Manby F~R and Sch{\"u}tz M 2012 {\em WIREs
  Comput. Mol. Sci.\/} {\bf 2} 242
  \urlprefix\url{https://doi.org/10.1002/wcms.82}

\bibitem{MOLPRO-JCP}
Werner H~J, Knowles P~J, Manby F~R, Black J~A, Doll K, He{\ss}elmann A, Kats D,
  K\"{o}hn A, Korona T, Kreplin D~A, Ma Q, {Miller III} T~F, Mitrushchenkov A,
  Peterson K~A, Polyak I, Rauhut G and Sibaev M 2020 {\em J. Chem. Phys.\/}
  {\bf 152} 144107 \urlprefix\url{https://doi.org/10.1063/5.0005081}

\bibitem{MOLPRO_brief}
Werner H~J, Knowles P~J {\em et~al.\/} Molpro, version 2021.1, a package of ab
  initio programs see https://www.molpro.net

\bibitem{Cizek:1966}
\v{C}\'{i}\v{z}ek J 1966 {\em J. Chem. Phys.\/} {\bf 45} 4256 ISSN 0021-9606
  \urlprefix\url{https://doi.org/10.1063/1.1727484}

\bibitem{Bartlett:1989}
Bartlett R~J 1989 {\em J. Phys. Chem.\/} {\bf 93} 1697
  \urlprefix\url{https://doi.org/10.1021/j100342a008}

\bibitem{Raghavachari_Head-Gordon:1989}
Raghavachari K, Trucks G~W, Pople J~A and Head-Gordon M 1989 {\em Chem. Phys.
  Lett.\/} {\bf 157} 479 ISSN 0009-2614
  \urlprefix\url{https://doi.org/10.1016/S0009-2614(89)87395-6}

\bibitem{Bartlett_Noga:1990}
Bartlett R~J, Watts J, Kucharski S and Noga J 1990 {\em Chem. Phys. Lett.\/}
  {\bf 165} 513--522 ISSN 0009-2614
  \urlprefix\url{https://doi.org/10.1016/0009-2614(90)87031-L}

\bibitem{Stanton:1997}
Stanton J~F 1997 {\em Chem. Phys. Lett.\/} {\bf 281} 130--134 ISSN 0009-2614
  \urlprefix\url{https://doi.org/10.1016/S0009-2614(97)01144-5}

\bibitem{Moller_Plesset:1934}
M\o{}ller C and Plesset M~S 1934 {\em Phys. Rev.\/} {\bf 46}(7) 618--622
  \urlprefix\url{https://doi.org/10.1103/PhysRev.46.618}

\bibitem{Cremer:2011}
Cremer D 2011 {\em WIREs Comput. Mol. Sci.\/} {\bf 1} 509
  \urlprefix\url{https://doi.org/10.1002/wcms.58}

\bibitem{Ma_Werner:2018_review}
Ma Q and Werner H~J 2018 {\em WIREs Comput. Mol. Sci.\/} {\bf 8} e1371
  \urlprefix\url{https://doi.org/10.1002/wcms.1371}

\bibitem{Zaspel:2018}
Zaspel P, Huang B, Harbrecht H and {von Lilienfeld} O~A 2018 {\em J. Chem.
  Theory Comput.\/} {\bf 15} 1546
  \urlprefix\url{https://doi.org/10.1021/acs.jctc.8b00832}

\bibitem{Heinen_Lilienfeld:2023}
Heinen S, Khan D, Falk~{von Rudorff} G, Karandashev K, Jose Arismendi~Arrieta
  D, Price A~J~A, Nandi S, Bhowmik A, Hermansson K and Anatole~{von Lilienfeld}
  O 2024 {\em Mach. Learn.: Sci. Technol.\/} {\bf 5} 025058
  \urlprefix\url{https://dx.doi.org/10.1088/2632-2153/ad4ae5}

\bibitem{cosmo}
Klamt A and Sch\"{u}\"{u}rmann G 1993 {\em J. Chem. Soc.{,} Perkin Trans. 2\/}
  (5) 799--805 \urlprefix\url{http://dx.doi.org/10.1039/P29930000799}

\bibitem{cosmo2}
Klamt A 1995 {\em J. Phys. Chem.\/} {\bf 99} 2224--2235
  \urlprefix\url{https://doi.org/10.1021/j100007a062}

\bibitem{KLAMT200043}
Klamt A and Eckert F 2000 {\em Fluid Phase Equilibria\/} {\bf 172} 43--72 ISSN
  0378-3812 \urlprefix\url{https://doi.org/10.1016/S0378-3812(00)00357-5}

\bibitem{doi:10.1021/ar800187p}
Klamt A, Mennucci B, Tomasi J, Barone V, Curutchet C, Orozco M and Luque F~J
  2009 {\em Acc. Chem. Research\/} {\bf 42} 489--492
  \urlprefix\url{https://doi.org/10.1021/ar800187p}

\bibitem{doi:10.1146/annurev-chembioeng-073009-100903}
Klamt A, Eckert F and Arlt W 2010 {\em Annu. Rev. Chem. Biomol. Eng.\/} {\bf 1}
  101--122
  \urlprefix\url{https://doi.org/10.1146/annurev-chembioeng-073009-100903}

\bibitem{Chung_Green:2022}
Chung Y, Vermeire F~H, Wu H, Walker P~J, Abraham M~H and Green W~H 2022 {\em J.
  Chem. Inf. Model\/} {\bf 62} 433--446 pMID: 35044781
  \urlprefix\url{https://doi.org/10.1021/acs.jcim.1c01103}

\bibitem{Weinreich_Lilienfeld:2022}
Weinreich J, Lemm D, {von Rudorff} G~F and {von Lilienfeld} O~A 2022 {\em J.
  Chem. Phys.\/} {\bf 157} 024303 ISSN 0021-9606
  \urlprefix\url{https://doi.org/10.1063/5.0095674}

\bibitem{doi:10.1021/acs.jcim.6b00081}
Zhang J, Tuguldur B and van~der Spoel D 2016 {\em J. Chem. Inf. Model.\/} {\bf
  56} 819--820 \urlprefix\url{https://doi.org/10.1021/acs.jcim.6b00081}

\bibitem{doi:10.1021/ie049139z}
Grensemann H and Gmehling J 2005 {\em Industrial \& Engineering Chemistry
  Research\/} {\bf 44} 1610--1624
  \urlprefix\url{https://doi.org/10.1021/ie049139z}

\bibitem{doi:10.1021/acs.jctc.2c00919}
Toman\'{i}k L, Rul\'{i}\v{s}ek L and Slav\'{i}\v{c}ek P 2023 {\em J. Chem.
  Theory Comput.\/} {\bf 19} 1014--1022
  \urlprefix\url{https://doi.org/10.1021/acs.jctc.2c00919}

\bibitem{Ruddigkeit_Reymond:2012}
Ruddigkeit L, {van Deursen} R, Blum L~C and Reymond J~L 2012 {\em J. Chem. Inf.
  Model.\/} {\bf 52} 2864 \urlprefix\url{https://doi.org/10.1021/ci300415d}

\bibitem{agz7}
Huang B and von Lilienfeld O~A 2020 Dictionary of 140k {GDB} and {ZINC} derived
  {AMON}s (\textit{Preprint} \eprint{2008.05260})

\bibitem{Ramakrishnan_Lilienfeld:2014}
Ramakrishnan R, Dral P~O, Rupp M and {von Lilienfeld} O~A 2014 {\em Sci.
  Data\/} {\bf 1} 140022 \urlprefix\url{https://doi.org/10.1038/sdata.2014.22}

\bibitem{Dunning:1989}
Dunning T~H 1989 {\em J. Chem. Phys.\/} {\bf 90} 1007
  \urlprefix\url{https://doi.org/10.1063/1.456153}

\bibitem{Kendall_Harrison:1992}
Kendall R~A, Dunning T~H and Harrison R~J 1992 {\em J. Chem. Phys.\/} {\bf 96}
  6796 \urlprefix\url{https://doi.org/10.1063/1.462569}

\bibitem{Woon_Dunning:1993}
Woon D~E and Dunning T~H 1993 {\em J. Chem. Phys.\/} {\bf 98} 1358--1371
  \urlprefix\url{https://doi.org/10.1063/1.464303}

\bibitem{Weigend_Ahlrichs:2005}
Weigend F and Ahlrichs R 2005 {\em Phys. Chem. Chem. Phys.\/} {\bf 7} 3297
  \urlprefix\url{https://doi.org/10.1039/b508541a}

\bibitem{Peterson_Dolg:2003}
Peterson K~A, Figgen D, Goll E, Stoll H and Dolg M 2003 {\em J. Chem. Phys.\/}
  {\bf 119} 11113--11123 \urlprefix\url{https://doi.org/10.1063/1.1622924}

\bibitem{Figgen_Stoll:2005}
Figgen D, Rauhut G, Dolg M and Stoll H 2005 {\em J. Chem. Phys.\/} {\bf 311}
  227--244 \urlprefix\url{https://doi.org/10.1016/j.chemphys.2004.10.005}

\bibitem{Peterson_Puzzarini:2005}
Peterson K~A and Puzzarini C 2005 {\em Theor. Chem. Acc.\/} {\bf 114} 283--296
  \urlprefix\url{https://doi.org/10.1007/s00214-005-0681-9}

\bibitem{Pulay_Saebo:1986}
Pulay P and Saeb\o{} S 1986 {\em Theor. Chim. Acta\/} {\bf 69} 357
  \urlprefix\url{https://doi.org/10.1007/BF00526697}

\bibitem{Hampel_Werner:1996}
Hampel C and Werner H 1996 {\em J. Chem. Phys.\/} {\bf 104} 6286--6297 ISSN
  0021-9606 \urlprefix\url{https://doi.org/10.1063/1.471289}

\bibitem{Schutz_Werner:2000}
Sch\"{u}tz M and Werner H~J 2000 {\em Chem. Phys. Lett.\/} {\bf 318} 370--378
  ISSN 0009-2614 \urlprefix\url{https://doi.org/10.1016/S0009-2614(00)00066-X}

\bibitem{Ahlrichs_Driessler:1975}
Ahlrichs R and Driessler F 1975 {\em Theor. Chim. Acta\/} {\bf 36} 275
  \urlprefix\url{https://doi.org/10.1007/BF00549691}

\bibitem{Taylor:1981}
Taylor P~R 1981 {\em J. Chem. Phys.\/} {\bf 74} 1256
  \urlprefix\url{https://doi.org/10.1063/1.441186}

\bibitem{Staemmler_Jaquet:1981}
Staemmler V and Jaquet R 1981 {\em Theor. Chim. Acta\/} {\bf 59} 129
  \urlprefix\url{https://doi.org/10.1007/BF00938691}

\bibitem{Riplinger_Neese:2013a}
Riplinger C and Neese F 2013 {\em J. Chem. Phys.\/} {\bf 138} 034106 ISSN
  0021-9606 \urlprefix\url{https://doi.org/10.1063/1.4773581}

\bibitem{Riplinger_Neese:2013b}
Riplinger C, Sandhoefer B, Hansen A and Neese F 2013 {\em J. Chem. Phys.\/}
  {\bf 139} 134101 ISSN 0021-9606
  \urlprefix\url{https://doi.org/10.1063/1.4821834}

\bibitem{Ten-no:2004a}
Ten-no S 2004 {\em J. Chem. Phys.\/} {\bf 121} 117--129
  \urlprefix\url{https://doi.org/10.1063/1.1757439}

\bibitem{Ten-no:2004b}
Ten-no S 2004 {\em Chem. Phys. Lett.\/} {\bf 398} 56
  \urlprefix\url{https://doi.org/10.1016/j.cplett.2004.09.041}

\bibitem{Adler_Werner:2007}
Adler T~B, Knizia G and Werner H~J 2007 {\em J. Chem. Phys.\/} {\bf 127} 221106
  \urlprefix\url{https://doi.org/10.1063/1.2817618}

\bibitem{Knizia_Werner:2009}
Knizia G, Adler T~B and Werner H~J 2009 {\em J. Chem. Phys.\/} {\bf 130} 054104
  ISSN 0021-9606 \urlprefix\url{https://doi.org/10.1063/1.3054300}

\bibitem{Knizia_Werner:2008}
Knizia G and Werner H~J 2008 {\em J. Chem. Phys.\/} {\bf 128} 154103
  \urlprefix\url{https://doi.org/10.1063/1.2889388}

\bibitem{Giner_Toulouse:2018}
Giner E, Pradines B, Fert\'{e} A, Assaraf R, Savin A and Toulouse J 2018 {\em
  J. Chem. Phys.\/} {\bf 149} 194301 ISSN 0021-9606
  \urlprefix\url{https://doi.org/10.1063/1.5052714}

\bibitem{Loos_Giner:2019}
Loos P~F, Pradines B, Scemama A, Toulouse J and Giner E 2019 {\em J. Phys.
  Chem. Lett.\/} {\bf 10} 2931 pMID: 31090432
  \urlprefix\url{https://doi.org/10.1021/acs.jpclett.9b01176}

\bibitem{Giner_Toulouse:2020}
Giner E, Scemama A, Loos P~F and Toulouse J 2020 {\em J. Chem. Phys.\/} {\bf
  152} 174104 ISSN 0021-9606 \urlprefix\url{https://doi.org/10.1063/5.0002892}

\bibitem{McNaught_Wilkinson:1997}
McNaught A and Wilkinson A 1997 {\em Compendium of chemical terminology\/} 2nd
  ed IUPAC Chemical Nomenclature S. (IUPAC International Union of Pure and
  Applied Chem)

\bibitem{Weininger:1988}
Weininger D 1988 {\em J. Chem. Inf. Comput. Sci.\/} {\bf 28} 31--36
  \urlprefix\url{https://doi.org/10.1021/ci00057a005}

\bibitem{Huang_Lilienfeld:2020}
Huang B and von Lilienfeld O~A 2020 {\em Nat. Chem.\/} {\bf 12}(10) 945
  \urlprefix\url{https://doi.org/10.1038/s41557-020-0527-z}

\bibitem{Tomasi_Cammi:2005}
Tomasi J, Mennucci B and Cammi R 2005 {\em Chem. Rev.\/} {\bf 105} 2999--3094
  pMID: 16092826 \urlprefix\url{https://doi.org/10.1021/cr9904009}

\bibitem{therm}
Eckert F and Klamt A 2018 {COSMOtherm} {BIOVIA} {COSMOtherm}, Release 2021;
  Dassault Syst\`{e}mes. http://www.3ds.com

\bibitem{TURBOMOLE}
{TURBOMOLE V7.2 2017}, a development of {University of Karlsruhe} and
  {Forschungszentrum Karlsruhe GmbH}, 1989-2007, {TURBOMOLE GmbH}, since 2007;
  available from \\ http://www.turbomole.com.

\bibitem{bp86}
Becke A~D 1988 {\em Phys. Rev. A\/} {\bf 38}(6) 3098--3100
  \urlprefix\url{https://doi.org/10.1103/PhysRevA.38.3098}

\bibitem{assesment}
Ahlrichs R, Furche F and Grimme S 2000 {\em Chem. Phys. Lett.\/} {\bf 325}
  317--321 \urlprefix\url{https://doi.org/10.1016/S0009-2614(00)00654-0}

\bibitem{metz2000a}
Metz B, Stoll H and Dolg M 2000 {\em J. Chem. Phys.\/} {\bf 113} 2563--2569
  \urlprefix\url{https://doi.org/10.1063/1.1305880}

\bibitem{rappoport2010a}
Rappoport D and Furche F 2010 {\em J. Chem. Phys.\/} {\bf 133} 134105
  \urlprefix\url{https://doi.org/10.1063/1.3484283}

\bibitem{COSMOconf_software}
{BIOVIA, Dassault Syst\`{e}mes} {COSMOconf}, Release 2021, San Diego: Dassault
  Syst\`{e}mes, 2021.

\bibitem{Hansen_Gunsteren:2014}
Hansen N and van Gunsteren W~F 2014 {\em J. Chem. Theory Comput.\/} {\bf 10}
  2632--2647 pMID: 26586503 \urlprefix\url{https://doi.org/10.1021/ct500161f}

\bibitem{Vainio_Johnson:2007}
Vainio M~J and Johnson M~S 2007 {\em J. Chem. Inf. Model.\/} {\bf 47} 2462
  \urlprefix\url{https://doi.org/10.1021/ci6005646}

\bibitem{Holland:1975}
Holland J~H 1975 {\em Adaptation in Natural and Artificial Systems\/}
  (University of Michigan Press, Ann Arbor)
  \urlprefix\url{https://doi.org/10.7551/mitpress/1090.001.0001}

\bibitem{Boyle_Dalke:2018}
O'Boyle N and Dalke A 2018 {\em ChemRxiv\/} 10.26434/chemrxiv.7097960.v1 pMID:
  26586503 \urlprefix\url{https://doi.org/10.26434/chemrxiv.7097960.v1}

\bibitem{Krenn_Aspuru-Guzik:2020}
Krenn M, H\"{a}se F, Nigam A, Friederich P and Aspuru-Guzik A 2020 {\em Mach.
  Learn.: Sci. Technol.\/} {\bf 1} 045024
  \urlprefix\url{https://dx.doi.org/10.1088/2632-2153/aba947}

\bibitem{Jensen:2007}
Jensen F 2007 {\em Introduction to Computational Chemistry\/} (John Wiley \&
  Sons Ltd) ISBN 978-0-470-01186-7

\bibitem{Montavon_Lilienfeld:2013}
Montavon G, Rupp M, Gobre V, Vazquez-Mayagoitia A, Hansen K, Tkatchenko A,
  M\"{u}ller K~R and {von Lilienfeld} O~A 2013 {\em New J. Phys.\/} {\bf 15}
  095003 \urlprefix\url{https://dx.doi.org/10.1088/1367-2630/15/9/095003}

\bibitem{Rogers_Hahn:2010}
Rogers D and Hahn M 2010 {\em J. Chem. Inf. Model.\/} {\bf 50} 742 pMID:
  20426451 \urlprefix\url{https://doi.org/10.1021/ci100050t}

\bibitem{software:RDKit}
  RDKit: Open-source cheminformatics. https://www.rdkit.org

\bibitem{zenodo_upload:2024}
Weinreich J, Karandashev K, Arismendi~Arrieta D~J, Hermansson K and von
  Lilienfeld A 2024 {Calculated state-of-the art results for solvation and
  ionization energies of thousands of organic molecules relevant to battery
  design} \urlprefix\url{https://doi.org/10.5281/zenodo.13952172}

\end{thebibliography}

\end{document}


\title[Calculated solvation and ionization energies]{Supplementary Data: Calculated solvation and ionization energies for thousands of organic molecules relevant to battery design}
\author{Jan Weinreich\footnote[2]{Ecole Polytechnique F\'{e}d\'{e}rale de Lausanne, Institut des Sciences et Ing\'{e}nierie chimiques, BCH 5312 (B\v{a}t. BCH)
CH-1015 Lausanne}\footnote[1]{Contributed equally to this work}, Konstantin Karandashev\footnote[3]{University of Vienna, Faculty of Physics, Kolingasse 14-16, AT-1090 Wien, Austria}\footnote[4]{\path{konstantin.karandashev@univie.ac.at}}$\dagger$, Daniel Jose Arismendi Arrieta\footnote[5]{{Department of Chemistry-{\AA}ngstr\"om Laboratory, Uppsala University Box 538, 75121, Uppsala, Sweden}}, Kersti Hermansson\footnotemark[5], O. Anatole von Lilienfeld\footnote[6]{Vector Institute for Artificial Intelligence, Toronto, ON, M5S 1M1, Canada}\footnote[7]{Departments of Chemistry, Materials Science and Engineering, and Physics, University of Toronto, St. George Campus, Toronto, ON, Canada}\footnote[8]{Machine Learning Group, Technische Universit\"at Berlin and Institute for the Foundations of Learning and Data, 10587 Berlin, Germany}}

\maketitle

\begin{abstract}

Contains additional tables and figures related to Section~4 (\textbf{Data overview}) of the main text.

\end{abstract}

\section{QM9-IPEA}

Tables~\ref{tab:QM9IPEA_extrema_Eatom}-\ref{tab:QM9IPEA_extrema_EA} present minimum and maximum values of atomization energy $\Eatom$, ionization energy $\IE$, and electron affinity $\EA$ calculated for QM9-IPEA with each method considered in this work. Interestingly, the post-HF methods mostly agree on which molecule exhibits the smallest or largest value of a given quantity.

\begin{table*}
 \centering
\begin{tabular}{llll}
\toprule
 \multirow{1}{*}{method} & \multirow{1}{*}{id} & \multirow{1}{*}{SMILES} & \multirow{1}{*}{$E_{\mathrm{atom}}$, kcal/mol} \\
\midrule \multicolumn{4}{c}{min.} \\
\midrule DF-HF & $121108$ & CCCC1CCC(C)C1 & $-2.04{ \cdot } 10^{3}$ \\
 DF-HF-CABS & $\phantom{0}86195$ & CC1CC(C)C(C)C1C & $-2.67{ \cdot } 10^{3}$ \\
 PNO-LMP2-F12 & $\phantom{0}86195$ & CC1CC(C)C(C)C1C & $-2.67{ \cdot } 10^{3}$ \\
 PNO-LCCSD-F12b & $121108$ & CCCC1CCC(C)C1 & $-2.60{ \cdot } 10^{3}$ \\
 PNO-LCCSD(T)-F12b & $121108$ & CCCC1CCC(C)C1 & $-2.64{ \cdot } 10^{3}$ \\
 PNO-LCCSD(T*)-F12b & $121108$ & CCCC1CCC(C)C1 & $-2.60{ \cdot } 10^{3}$ \\
\midrule \multicolumn{4}{c}{max.} \\
\midrule DF-HF & $\phantom{0}\phantom{0}3839$ & N=C1N=NNN=N1 & $-3.89{ \cdot } 10^{2}$ \\
 DF-HF-CABS & $\phantom{0}\phantom{0}\phantom{0}827$ & FC(F)(F)C\#N & $-6.78{ \cdot } 10^{2}$ \\
 PNO-LMP2-F12 & $\phantom{0}\phantom{0}\phantom{0}827$ & FC(F)(F)C\#N & $-6.80{ \cdot } 10^{2}$ \\
 PNO-LCCSD-F12b & $\phantom{0}\phantom{0}\phantom{0}827$ & FC(F)(F)C\#N & $-6.19{ \cdot } 10^{2}$ \\
 PNO-LCCSD(T)-F12b & $\phantom{0}\phantom{0}\phantom{0}827$ & FC(F)(F)C\#N & $-6.39{ \cdot } 10^{2}$ \\
 PNO-LCCSD(T*)-F12b & $\phantom{0}\phantom{0}\phantom{0}827$ & FC(F)(F)C\#N & $-6.17{ \cdot } 10^{2}$ \\
\bottomrule
\end{tabular}

\caption{Lowest ("min.") and highest ("max.") values of atomization energy $\Eatom$ in QM9-IPEA for each method, along with the QM9 index ("id") of the molecule for which it was observed and the latter's SMILES.}
\label{tab:QM9IPEA_extrema_Eatom}
\end{table*}

 \begin{table*}
 \centering
\begin{tabular}{llll}
\toprule
 \multirow{1}{*}{method} & \multirow{1}{*}{id} & \multirow{1}{*}{SMILES} & \multirow{1}{*}{IE, eV} \\
\midrule \multicolumn{4}{c}{min.} \\
\midrule DF-HF & $\phantom{0}28102$ & CC1=CC2=C(N1)C=CC2 & $\phantom{0}5.88$ \\
 DF-HF-CABS & $101807$ & {[}NH3+{]}CC(O)CCC({[}O-{]})=O & $\phantom{0}6.69$ \\
 PNO-LMP2-F12 & $101807$ & {[}NH3+{]}CC(O)CCC({[}O-{]})=O & $\phantom{0}6.71$ \\
 PNO-LCCSD-F12b & $\phantom{0}24464$ & N1C=CC2=CC=CC2=C1 & $\phantom{0}6.92$ \\
 PNO-LCCSD(T)-F12b & $101807$ & {[}NH3+{]}CC(O)CCC({[}O-{]})=O & $\phantom{0}6.85$ \\
 PNO-LCCSD(T*)-F12b & $\phantom{0}24464$ & N1C=CC2=CC=CC2=C1 & $\phantom{0}6.91$ \\
\midrule \multicolumn{4}{c}{max.} \\
\midrule DF-HF & $\phantom{0}99424$ & COC(C\#N)(C\#N)C\#N & $13.27$ \\
 DF-HF-CABS & $\phantom{0}\phantom{0}\phantom{0}827$ & FC(F)(F)C\#N & $14.76$ \\
 PNO-LMP2-F12 & $\phantom{0}\phantom{0}\phantom{0}827$ & FC(F)(F)C\#N & $14.76$ \\
 PNO-LCCSD-F12b & $\phantom{0}\phantom{0}\phantom{0}827$ & FC(F)(F)C\#N & $14.21$ \\
 PNO-LCCSD(T)-F12b & $\phantom{0}\phantom{0}\phantom{0}827$ & FC(F)(F)C\#N & $14.33$ \\
 PNO-LCCSD(T*)-F12b & $\phantom{0}\phantom{0}\phantom{0}827$ & FC(F)(F)C\#N & $14.21$ \\
\bottomrule
\end{tabular}

\caption{Lowest and highest values of ionization energy $\IE$ in QM9-IPEA for each method, fields labeled analogously to Table~\ref{tab:QM9IPEA_extrema_Eatom}.}
\label{tab:QM9IPEA_extrema_IE}
\end{table*}

 \begin{table*}
 \centering
\begin{tabular}{llll}
\toprule
 \multirow{1}{*}{method} & \multirow{1}{*}{id} & \multirow{1}{*}{SMILES} & \multirow{1}{*}{EA, eV} \\
\midrule \multicolumn{4}{c}{min.} \\
\midrule DF-HF & $\phantom{0}\phantom{0}\phantom{0}827$ & FC(F)(F)C\#N & $-1.23\phantom{{ \cdot } 10^{0\phantom{-}}}$ \\
 DF-HF-CABS & $\phantom{0}\phantom{0}\phantom{0}827$ & FC(F)(F)C\#N & $-9.42{ \cdot } 10^{-1}$ \\
 PNO-LMP2-F12 & $\phantom{0}\phantom{0}\phantom{0}827$ & FC(F)(F)C\#N & $-9.37{ \cdot } 10^{-1}$ \\
 PNO-LCCSD-F12b & $\phantom{0}\phantom{0}\phantom{0}827$ & FC(F)(F)C\#N & $-8.85{ \cdot } 10^{-1}$ \\
 PNO-LCCSD(T)-F12b & $\phantom{0}\phantom{0}\phantom{0}827$ & FC(F)(F)C\#N & $-8.57{ \cdot } 10^{-1}$ \\
 PNO-LCCSD(T*)-F12b & $\phantom{0}\phantom{0}\phantom{0}827$ & FC(F)(F)C\#N & $-8.89{ \cdot } 10^{-1}$ \\
\midrule \multicolumn{4}{c}{max.} \\
\midrule DF-HF & $\phantom{0}23832$ & FC1=NC(=O)ON=N1 & $\phantom{-}1.39\phantom{{ \cdot } 10^{0\phantom{-}}}$ \\
 DF-HF-CABS & $130729$ & O=C1NN=NC(=N1)C\#N & $\phantom{-}2.06\phantom{{ \cdot } 10^{0\phantom{-}}}$ \\
 PNO-LMP2-F12 & $130729$ & O=C1NN=NC(=N1)C\#N & $\phantom{-}2.06\phantom{{ \cdot } 10^{0\phantom{-}}}$ \\
 PNO-LCCSD-F12b & $130729$ & O=C1NN=NC(=N1)C\#N & $\phantom{-}2.03\phantom{{ \cdot } 10^{0\phantom{-}}}$ \\
 PNO-LCCSD(T)-F12b & $130729$ & O=C1NN=NC(=N1)C\#N & $\phantom{-}2.00\phantom{{ \cdot } 10^{0\phantom{-}}}$ \\
 PNO-LCCSD(T*)-F12b & $130729$ & O=C1NN=NC(=N1)C\#N & $\phantom{-}2.03\phantom{{ \cdot } 10^{0\phantom{-}}}$ \\
\bottomrule
\end{tabular}

\caption{Lowest and highest values of electron affinity $\EA$ in QM9-IPEA for each method, fields labeled analogously to Table~\ref{tab:QM9IPEA_extrema_Eatom}.}
\label{tab:QM9IPEA_extrema_EA}
\end{table*}

\begin{figure*}
\center
\includegraphics[width=1.\textwidth]{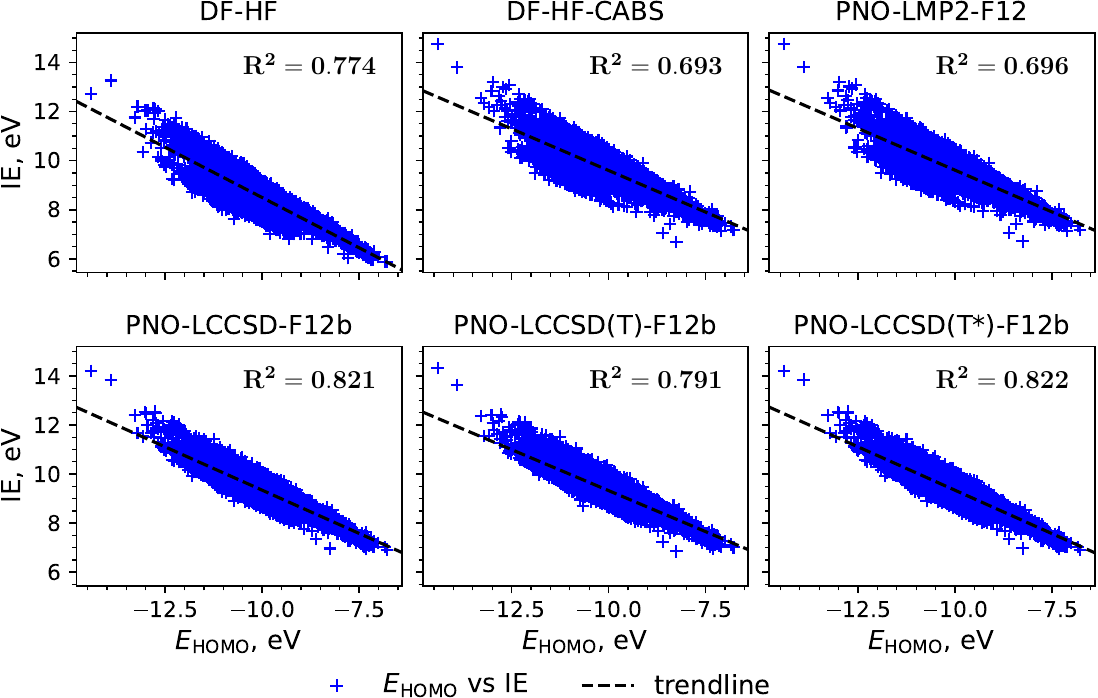}
    \caption{Values of HOMO energy $\HOMO$ vs. ionization energy $\IE$ plotted for all QM9-IPEA calculations for different methods considered in this work. Also shown are the linear trendlines and their $R^{2}$ factors for predicting $\IE$ as a function of $\HOMO$.}
    \label{fig:QM9IPEA_Koopman_HOMO} 
\end{figure*}

\begin{figure*}
\center
\includegraphics[width=1.\textwidth]{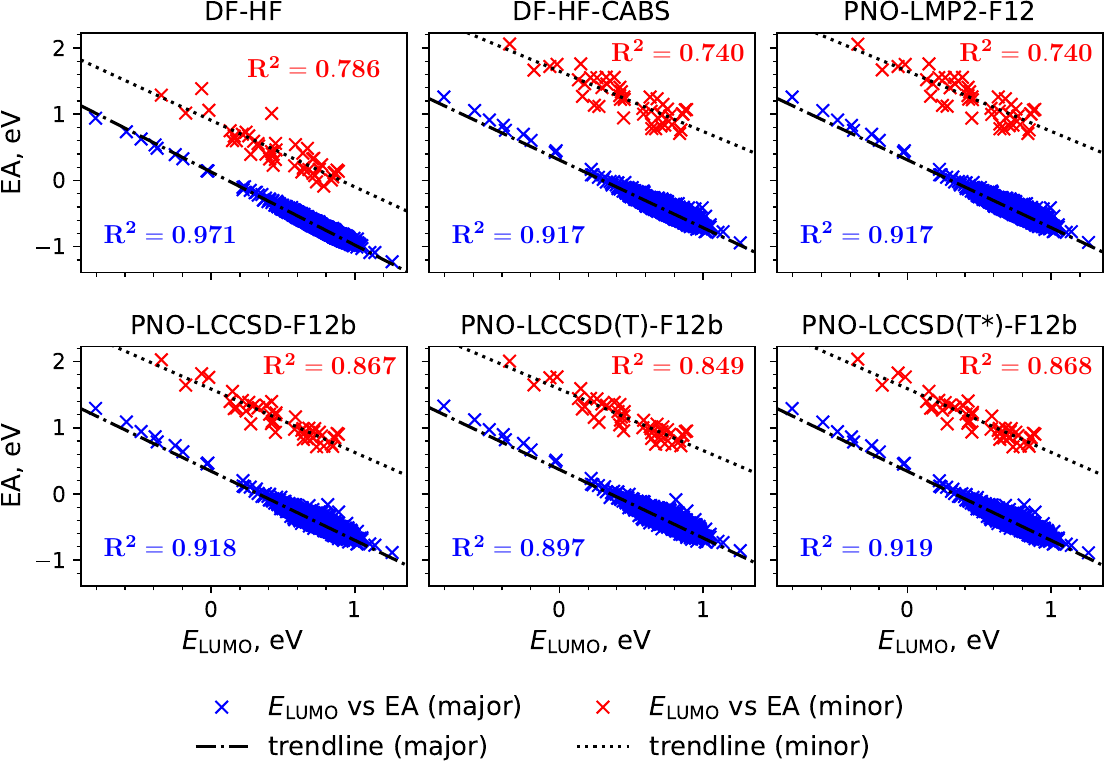}
          \caption{Values of LUMO energy $\LUMO$ vs. electron affinity $\EA$ plotted for all QM9-IPEA calculations for different methods considered in this work; the molecules are split into ``major'' and ``minor'' subsets as explained in Subsec.~4.2 of the main text. Also shown are the linear trendlines for both sets and their $R^{2}$ factors for predicting $\EA$ as a function of $\LUMO$.}
     \label{fig:QM9IPEA_Koopman_LUMO} 
\end{figure*}

Figures~\ref{fig:QM9IPEA_Koopman_HOMO} and~\ref{fig:QM9IPEA_Koopman_LUMO} present correlations between $\IE$ and $\EA$ with HOMO and LUMO energies ($\HOMO$ and $\LUMO$); the corresponding linear fit formulas and their $R^{2}$ factors are gathered in Tables~\ref{tab:HOMO_Koopman} and~\ref{tab:LUMO_Koopman}. Molecules constituting the "minor" subset of QM9-IPEA (as defined in the main text) are listed in Table~\ref{tab:weird_LUMO}.

\begin{table}
\begin{tabular}{lccc}
\toprule
 method & slope & intercept & $R^{2}$ \\
\midrule DF-HF & $-0.819$ & $0.305$ & $0.774$ \\
 DF-HF-CABS & $-0.679$ & $2.807$ & $0.693$ \\
 PNO-LMP2-F12 & $-0.681$ & $2.799$ & $0.696$ \\
 PNO-LCCSD-F12b & $-0.705$ & $2.289$ & $0.821$ \\
 PNO-LCCSD(T)-F12b & $-0.669$ & $2.641$ & $0.791$ \\
 PNO-LCCSD(T*)-F12b & $-0.708$ & $2.265$ & $0.822$ \\
\bottomrule
\end{tabular}

\caption{Slopes, intercepts, and $R^{2}$ factors observed while fitting ionization energy as a linear function of HOMO energy for different methods considered in this work.}
\label{tab:HOMO_Koopman}
\end{table}

\begin{table}
\begin{tabular}{lccccccc}
\toprule
 \multirow{2}{*}{method} & \multicolumn{3}{c}{major} & \phantom{\_} & \multicolumn{3}{c}{minor} \\
\cline{2-4}\cline{6-8}  & slope & intercept & $R^{2}$ & \phantom{\_} & slope & intercept & $R^{2}$ \\
\midrule DF-HF & $-1.103$ & $0.127$ & $0.971$ & \phantom{\_} & $-1.005$ & $0.907$ & $0.786$ \\
 DF-HF-CABS & $-1.022$ & $0.308$ & $0.917$ & \phantom{\_} & $-0.908$ & $1.648$ & $0.740$ \\
 PNO-LMP2-F12 & $-1.022$ & $0.308$ & $0.917$ & \phantom{\_} & $-0.907$ & $1.648$ & $0.740$ \\
 PNO-LCCSD-F12b & $-1.042$ & $0.342$ & $0.918$ & \phantom{\_} & $-0.957$ & $1.583$ & $0.867$ \\
 PNO-LCCSD(T)-F12b & $-1.032$ & $0.367$ & $0.897$ & \phantom{\_} & $-0.928$ & $1.583$ & $0.849$ \\
 PNO-LCCSD(T*)-F12b & $-1.043$ & $0.340$ & $0.919$ & \phantom{\_} & $-0.960$ & $1.588$ & $0.868$ \\
\bottomrule
\end{tabular}

\caption{Slopes, intercepts, and $R^{2}$ factors observed while fitting electron affinity as a linear function of LUMO energy for different methods considered in this work. Separate values are presented for the ``major'' and ``minor'' subsets which are defined in Subsec.~4.2 of the main text.}
\label{tab:LUMO_Koopman}
\end{table}

\begin{table}
\begin{tabular}{ll|ll}
\toprule
 id & SMILES & id & SMILES \\
\midrule $\phantom{0}\phantom{0}1585$ & NC(=N)C(=O)C\#N & $123514$ & C1=CN2N=NC=NC2=N1 \\
 $\phantom{0}\phantom{0}9117$ & CCC\#CC(=O)C\#N & $124262$ & CC(=O)C1=CN=NN=N1 \\
 $\phantom{0}21429$ & N=C1OC(=O)ON=N1 & $128861$ & N\#CC\#CC1=CN=NO1 \\
 $\phantom{0}21783$ & O=CC1=CN=NC=N1 & $128884$ & N\#CC1=NC(=NO1)C\#N \\
 $\phantom{0}22029$ & ON=C1C=CN=NC=C1 & $129100$ & N=C1C=NON=NC1=O$\dagger$ \\
 $\phantom{0}23832$ & FC1=NC(=O)ON=N1 & $129233$ & N=C1NN=NC=C1C=O \\
 $\phantom{0}24741$ & O=C(C\#C)C1=NC=CO1 & $129261$ & N=C1ON=C(C=O)C=C1 \\
 $\phantom{0}24767$ & O=C1C=CC2=C1OC=N2 & $129294$ & N=C1ON=CC=C1C=O \\
 $\phantom{0}25151$ & O=C1OC=NC(=C1)C\#C & $130525$ & O1N=NC2=C1C=NC=N2 \\
 $\phantom{0}25289$ & O=CC\#CC1=NC=CO1 & $130538$ & O1N=NC=C2N=CN=C12$\dagger$ \\
 $\phantom{0}25394$ & O=CC1=CN=CC(=O)N1 & $130566$ & O=C(C\#N)C1=CON=N1 \\
 $\phantom{0}25408$ & O=CC1=CN=COC1=O & $130729$ & O=C1NN=NC(=N1)C\#N \\
 $\phantom{0}27459$ & OC1=CC(=O)C=CN=C1 & $130734$ & O=C1NN=NC=CC1=O \\
 $\phantom{0}42353$ & O=C(CC\#N)C\#CC\#N & $130829$ & O=CC(=O)C1=NON=C1 \\
 $\phantom{0}43433$ & N=C1OC(=O)C(=C1)C\#C & $130841$ & O=CC1=C(C=NO1)C\#C \\
 $\phantom{0}49135$ & O=CC1=CC(=O)COC1 & $130848$ & O=CC1=C(N=NO1)C\#N \\
 $\phantom{0}52387$ & O=CCOCC(=O)C=O & $130883$ & O=CC1=NC(=O)N=NO1$\dagger$ \\
 $\phantom{0}60524$ & OC(C1CO1)C(=O)C=O & $130905$ & O=CC1=NOC(=C1)C\#N \\
 $\phantom{0}66498$ & OC12C3C1C1=CC2N3C1 & $131020$ & O=CN1N=CC(=N1)C\#N \\
 $\phantom{0}85350$ & OC1CC(=O)C=C1C\#N & $131041$ & O=CN1N=NC=NC1=O \\
 $\phantom{0}97529$ & {[}NH3+{]}CC\#C{[}C-{]}(C\#C)C\#N$\dagger$ & $131977$ & NC1=C(ON=N1)N(=O)=O \\
 $116844$ & OCCC\#CC(=O)C=O & $133188$ & CC1=NC(=O)ON=C1F \\
 $117036$ & COCC(=O)C(=O)C=O & $133384$ & NC1=NOC(=O)C(F)=N1 \\
\bottomrule
\end{tabular}

\caption{QM9 indices (``id'') and SMILES of molecules constituting the ``minor'' subset of QM9-IPEA. The ``$\dagger$'' symbol marks molecules whose QM9 SMILES is inaccurate due to atom reorganization during coordinate optimization.}
\label{tab:weird_LUMO}
\end{table}

Lastly, Tables~\ref{tab:atomization_energy_correlations}, \ref{tab:ionization_energy_correlations}, and~\ref{tab:electron_affinity_correlations} summarize $R^{2}$ scores for fitting $\IE$, $\EA$, and $\Eatom$ values as a linear function of the quantity at a different level of theory, with the corresponding linear fit coefficients presented in Tables~\ref{tab:atomization_energy_correlations_formulas}, \ref{tab:ionization_energy_correlations_formulas}, and~\ref{tab:electron_affinity_correlations_formulas}. Very tight correlation between all Coupled Cluster based methods considered in this work may mean large cancellation of rescaled triples excitation correction when $\IE$, $\EA$, and $\Eatom$ are calculated. We also note that high $R^{2}$ values for linearly fitting $\Eatom$ might be related to $\Eatom$ being an extensive quantity that can be rather accurately decomposed into atomic contributions (the corresponding $R^{2}$ score is $0.985$ for DF-HF and larger than $0.99$ for all other methods).

\begin{table}
\begin{tabular}{lcccccc}
\toprule
 \multirow{2}{*}{fitted method} & \multicolumn{6}{c}{linear fit's $R^{2}$ score} \\
\cline{2-7}  & \rotatebox[origin=c]{75}{DF-HF} & \rotatebox[origin=c]{75}{DF-HF-CABS} & \rotatebox[origin=c]{75}{PNO-LMP2-F12} & \rotatebox[origin=c]{75}{PNO-LCCSD-F12b} & \rotatebox[origin=c]{75}{PNO-LCCSD(T)-F12b} & \rotatebox[origin=c]{75}{PNO-LCCSD(T*)-F12b} \\
\midrule DF-HF & $1.00$ & $0.99$ & $0.99$ & $0.99$ & $0.99$ & $0.99$ \\
 DF-HF-CABS & $0.99$ & $1.00$ & $1.00$ & $1.00$ & $1.00$ & $1.00$ \\
 PNO-LMP2-F12 & $0.99$ & $1.00$ & $1.00$ & $1.00$ & $1.00$ & $1.00$ \\
 PNO-LCCSD-F12b & $0.99$ & $1.00$ & $1.00$ & $1.00$ & $1.00$ & $1.00$ \\
 PNO-LCCSD(T)-F12b & $0.99$ & $1.00$ & $1.00$ & $1.00$ & $1.00$ & $1.00$ \\
 PNO-LCCSD(T*)-F12b & $0.99$ & $1.00$ & $1.00$ & $1.00$ & $1.00$ & $1.00$ \\
\bottomrule
\end{tabular}

\caption{\Rvals{atomization energy}{atomization energy}}
\label{tab:atomization_energy_correlations}
\end{table}

\begin{table}
\begin{tabular}{lcccccc}
\toprule
 \multirow{2}{*}{fitted method} & \multicolumn{6}{c}{linear fit's slope and intercept} \\
\cline{2-7}  & \rotatebox[origin=c]{75}{DF-HF} & \rotatebox[origin=c]{75}{DF-HF-CABS} & \rotatebox[origin=c]{75}{PNO-LMP2-F12} & \rotatebox[origin=c]{75}{PNO-LCCSD-F12b} & \rotatebox[origin=c]{75}{PNO-LCCSD(T)-F12b} & \rotatebox[origin=c]{75}{PNO-LCCSD(T*)-F12b} \\
\midrule DF-HF & 1.00,$\phantom{-}$0.00 & 0.89,$\phantom{-}0.55$ & 0.89,$\phantom{-}0.56$ & 0.87,$\phantom{-}0.39$ & 0.88,$\phantom{-}0.45$ & 0.87,$\phantom{-}0.39$ \\
 DF-HF-CABS & 1.11,$-0.66$ & 1.00,$\phantom{-}$0.00 & 1.00,$\phantom{-}0.00$ & 0.98,$-0.19$ & 0.99,$-0.12$ & 0.98,$-0.20$ \\
 PNO-LMP2-F12 & 1.11,$-0.66$ & 1.00,$-0.00$ & 1.00,$\phantom{-}$0.00 & 0.98,$-0.19$ & 0.99,$-0.13$ & 0.98,$-0.20$ \\
 PNO-LCCSD-F12b & 1.13,$-0.47$ & 1.02,$\phantom{-}0.19$ & 1.02,$\phantom{-}0.19$ & 1.00,$\phantom{-}$0.00 & 1.00,$\phantom{-}0.07$ & 1.00,$-0.01$ \\
 PNO-LCCSD(T)-F12b & 1.13,$-0.54$ & 1.01,$\phantom{-}0.12$ & 1.01,$\phantom{-}0.13$ & 1.00,$-0.07$ & 1.00,$\phantom{-}$0.00 & 1.00,$-0.07$ \\
 PNO-LCCSD(T*)-F12b & 1.13,$-0.47$ & 1.02,$\phantom{-}0.20$ & 1.02,$\phantom{-}0.20$ & 1.00,$\phantom{-}0.01$ & 1.00,$\phantom{-}0.07$ & 1.00,$\phantom{-}$0.00 \\
\bottomrule
\end{tabular}

\caption{\slopeinter{atomization energy}{atomization energy}}
\label{tab:atomization_energy_correlations_formulas}
\end{table}

\begin{table}
\begin{tabular}{lcccccc}
\toprule
 \multirow{2}{*}{fitted method} & \multicolumn{6}{c}{linear fit's $R^{2}$ score} \\
\cline{2-7}  & \rotatebox[origin=c]{75}{DF-HF} & \rotatebox[origin=c]{75}{DF-HF-CABS} & \rotatebox[origin=c]{75}{PNO-LMP2-F12} & \rotatebox[origin=c]{75}{PNO-LCCSD-F12b} & \rotatebox[origin=c]{75}{PNO-LCCSD(T)-F12b} & \rotatebox[origin=c]{75}{PNO-LCCSD(T*)-F12b} \\
\midrule DF-HF & $1.00$ & $0.55$ & $0.55$ & $0.84$ & $0.77$ & $0.85$ \\
 DF-HF-CABS & $0.55$ & $1.00$ & $1.00$ & $0.87$ & $0.90$ & $0.87$ \\
 PNO-LMP2-F12 & $0.55$ & $1.00$ & $1.00$ & $0.87$ & $0.90$ & $0.87$ \\
 PNO-LCCSD-F12b & $0.84$ & $0.87$ & $0.87$ & $1.00$ & $0.99$ & $1.00$ \\
 PNO-LCCSD(T)-F12b & $0.77$ & $0.90$ & $0.90$ & $0.99$ & $1.00$ & $0.98$ \\
 PNO-LCCSD(T*)-F12b & $0.85$ & $0.87$ & $0.87$ & $1.00$ & $0.98$ & $1.00$ \\
\bottomrule
\end{tabular}

\caption{\Rvals{ionization energy}{ionization energy}}
\label{tab:ionization_energy_correlations}
\end{table}

\begin{table}
\begin{tabular}{lcccccc}
\toprule
 \multirow{2}{*}{fitted method} & \multicolumn{6}{c}{linear fit's slope and intercept} \\
\cline{2-7}  & \rotatebox[origin=c]{75}{DF-HF} & \rotatebox[origin=c]{75}{DF-HF-CABS} & \rotatebox[origin=c]{75}{PNO-LMP2-F12} & \rotatebox[origin=c]{75}{PNO-LCCSD-F12b} & \rotatebox[origin=c]{75}{PNO-LCCSD(T)-F12b} & \rotatebox[origin=c]{75}{PNO-LCCSD(T*)-F12b} \\
\midrule DF-HF & 1.00,$\phantom{-}$0.00 & 0.85,$\phantom{-}0.01$ & 0.85,$\phantom{-}0.01$ & 1.10,$-0.06$ & 1.08,$-0.06$ & 1.10,$-0.06$ \\
 DF-HF-CABS & 0.65,$\phantom{-}0.15$ & 1.00,$\phantom{-}$0.00 & 1.00,$-0.00$ & 0.98,$\phantom{-}0.02$ & 1.03,$-0.00$ & 0.97,$\phantom{-}0.02$ \\
 PNO-LMP2-F12 & 0.65,$\phantom{-}0.15$ & 1.00,$\phantom{-}0.00$ & 1.00,$\phantom{-}$0.00 & 0.98,$\phantom{-}0.02$ & 1.03,$-0.00$ & 0.97,$\phantom{-}0.02$ \\
 PNO-LCCSD-F12b & 0.77,$\phantom{-}0.10$ & 0.89,$\phantom{-}0.03$ & 0.89,$\phantom{-}0.03$ & 1.00,$\phantom{-}$0.00 & 1.03,$-0.01$ & 1.00,$\phantom{-}0.00$ \\
 PNO-LCCSD(T)-F12b & 0.71,$\phantom{-}0.12$ & 0.88,$\phantom{-}0.03$ & 0.88,$\phantom{-}0.03$ & 0.96,$\phantom{-}0.01$ & 1.00,$\phantom{-}$0.00 & 0.96,$\phantom{-}0.01$ \\
 PNO-LCCSD(T*)-F12b & 0.77,$\phantom{-}0.10$ & 0.89,$\phantom{-}0.03$ & 0.89,$\phantom{-}0.03$ & 1.00,$-0.00$ & 1.03,$-0.01$ & 1.00,$\phantom{-}$0.00 \\
\bottomrule
\end{tabular}

\caption{\slopeinter{ionization energy}{ionization energy}}
\label{tab:ionization_energy_correlations_formulas}
\end{table}

\begin{table}
\begin{tabular}{lcccccc}
\toprule
 \multirow{2}{*}{fitted method} & \multicolumn{6}{c}{linear fit's $R^{2}$ score} \\
\cline{2-7}  & \rotatebox[origin=c]{75}{DF-HF} & \rotatebox[origin=c]{75}{DF-HF-CABS} & \rotatebox[origin=c]{75}{PNO-LMP2-F12} & \rotatebox[origin=c]{75}{PNO-LCCSD-F12b} & \rotatebox[origin=c]{75}{PNO-LCCSD(T)-F12b} & \rotatebox[origin=c]{75}{PNO-LCCSD(T*)-F12b} \\
\midrule DF-HF & $1.00$ & $0.91$ & $0.91$ & $0.93$ & $0.92$ & $0.93$ \\
 DF-HF-CABS & $0.91$ & $1.00$ & $1.00$ & $0.99$ & $0.99$ & $0.99$ \\
 PNO-LMP2-F12 & $0.91$ & $1.00$ & $1.00$ & $0.99$ & $0.99$ & $0.99$ \\
 PNO-LCCSD-F12b & $0.93$ & $0.99$ & $0.99$ & $1.00$ & $1.00$ & $1.00$ \\
 PNO-LCCSD(T)-F12b & $0.92$ & $0.99$ & $0.99$ & $1.00$ & $1.00$ & $1.00$ \\
 PNO-LCCSD(T*)-F12b & $0.93$ & $0.99$ & $0.99$ & $1.00$ & $1.00$ & $1.00$ \\
\bottomrule
\end{tabular}

\caption{\Rvals{electron affinity}{electron affinity}}
\label{tab:electron_affinity_correlations}
\end{table}

\begin{table}
\begin{tabular}{lcccccc}
\toprule
 \multirow{2}{*}{fitted method} & \multicolumn{6}{c}{linear fit's slope and intercept} \\
\cline{2-7}  & \rotatebox[origin=c]{75}{DF-HF} & \rotatebox[origin=c]{75}{DF-HF-CABS} & \rotatebox[origin=c]{75}{PNO-LMP2-F12} & \rotatebox[origin=c]{75}{PNO-LCCSD-F12b} & \rotatebox[origin=c]{75}{PNO-LCCSD(T)-F12b} & \rotatebox[origin=c]{75}{PNO-LCCSD(T*)-F12b} \\
\midrule DF-HF & 1.00,$\phantom{-}$0.00 & 0.80,$-0.01$ & 0.80,$-0.01$ & 0.83,$-0.01$ & 0.83,$-0.01$ & 0.83,$-0.01$ \\
 DF-HF-CABS & 1.15,$\phantom{-}0.01$ & 1.00,$\phantom{-}$0.00 & 1.00,$-0.00$ & 1.03,$-0.00$ & 1.04,$-0.00$ & 1.03,$-0.00$ \\
 PNO-LMP2-F12 & 1.15,$\phantom{-}0.01$ & 1.00,$\phantom{-}0.00$ & 1.00,$\phantom{-}$0.00 & 1.03,$-0.00$ & 1.04,$-0.00$ & 1.03,$-0.00$ \\
 PNO-LCCSD-F12b & 1.12,$\phantom{-}0.01$ & 0.96,$-0.00$ & 0.96,$-0.00$ & 1.00,$\phantom{-}$0.00 & 1.01,$-0.00$ & 1.00,$\phantom{-}0.00$ \\
 PNO-LCCSD(T)-F12b & 1.11,$\phantom{-}0.01$ & 0.95,$\phantom{-}0.00$ & 0.95,$\phantom{-}0.00$ & 0.99,$\phantom{-}0.00$ & 1.00,$\phantom{-}$0.00 & 0.99,$\phantom{-}0.00$ \\
 PNO-LCCSD(T*)-F12b & 1.12,$\phantom{-}0.01$ & 0.96,$-0.00$ & 0.96,$-0.00$ & 1.00,$-0.00$ & 1.01,$-0.00$ & 1.00,$\phantom{-}$0.00 \\
\bottomrule
\end{tabular}

\caption{\slopeinter{electron affinity}{electron affinity}}
\label{tab:electron_affinity_correlations_formulas}
\end{table}

\section{SolQuest}

Keywords corresponding to different solvents for which solvation energies are published are listed in Table~\ref{tab:available_solvents}. Tables~\ref{tab:SolQuest_extrema_solvation} and~\ref{tab:SolQuest_extrema_total_energy} present minimum and maximum values of solvation energies and total energy (see explanation of \jsonkeyword{ENERGY} keyword in Subsec.~4.2 of the main text for definition) for SolQuest molecules.

\begin{table}
\begin{tabular}{lll}
\toprule
 \jsonkeyword{1,2-dichloroethane} & \jsonkeyword{ch2cl2} & \jsonkeyword{h2o} \\
 \jsonkeyword{1,2-dimethylbenzene} & \jsonkeyword{chcl3} & \jsonkeyword{hexamethylphosphoramide} \\
 \jsonkeyword{1,3-dimethylbenzene} & \jsonkeyword{chlorobenzene} & \jsonkeyword{hexane} \\
 \jsonkeyword{1,4-dimethylbenzene} & \jsonkeyword{cyclohexane} & \jsonkeyword{methanol} \\
 \jsonkeyword{1-butanol} & \jsonkeyword{diethyleneglycol} & \jsonkeyword{n-heptane} \\
 \jsonkeyword{1-octanol} & \jsonkeyword{diethylether} & \jsonkeyword{nitromethane} \\
 \jsonkeyword{2-butanol} & \jsonkeyword{diglyme} & \jsonkeyword{pentane} \\
 \jsonkeyword{2-propanol} & \jsonkeyword{dimethylsulfoxide} & \jsonkeyword{propanol} \\
 \jsonkeyword{aceticacid} & \jsonkeyword{dioxane} & \jsonkeyword{propanone} \\
 \jsonkeyword{acetonitrile} & \jsonkeyword{ethanol} & \jsonkeyword{pyridine} \\
 \jsonkeyword{benzene} & \jsonkeyword{ethylacetate} & \jsonkeyword{thf} \\
 \jsonkeyword{butanone} & \jsonkeyword{glycerol} & \jsonkeyword{toluene} \\
 \jsonkeyword{ccl4} & \jsonkeyword{glycol} & \jsonkeyword{triethylamine} \\
\bottomrule
\end{tabular}

\caption{SolQuest keywords corresponding to solvents for which solvation energies are published.}
\label{tab:available_solvents}
\end{table}

\begin{table*}
 \centering
\begin{tabular}{llll}
\toprule
 \multirow{1}{*}{source} & \multirow{2}{*}{solvent} & \multirow{2}{*}{SMILES} & \multirow{2}{*}{$E_{\mathrm{solv}}$, kcal/mol} \\
 \multirow{1}{*}{dataset} &  &  &  \\
\midrule \multicolumn{4}{c}{min} \\
\midrule \multirow{3}{*}{\begin{tabular}[t]{@{}c@{}}GDB17\\amons\end{tabular}} & pentane & NC(={[}NH2+{]})S(=O)(=O){[}O-{]} & $-1.10{ \cdot } 10^{1}$ \\
  & acetonitrile & NC(={[}NH2+{]})S(=O)(=O){[}O-{]} & $-2.97{ \cdot } 10^{1}$ \\
  & water & NCCS(=O)(=O)O & $-3.55{ \cdot } 10^{1}$ \\
\cline{1-2} \multirow{3}{*}{\begin{tabular}[t]{@{}c@{}}ZINC\\amons\end{tabular}} & pentane & CN(C)C={[}NH2+{]}.{[}Cl-{]} & $-1.21{ \cdot } 10^{1}$ \\
  & acetonitrile & NC(={[}NH2+{]}){[}C@H{]}(N)Cl.{[}Cl-{]} & $-3.07{ \cdot } 10^{1}$ \\
  & water & NC1={[}NH+{]}CP({[}O-{]})C1 & $-3.59{ \cdot } 10^{1}$ \\
\cline{1-2} \multirow{6}{*}{GDB17} & pentane & \begin{tabular}[t]{@{}l@{}}C{[}C@@{]}12{[}C@@H{]}(C\#N)C{[}C@@H{]}(C/N=C{\textbackslash}{[}NH\\-{]}){[}C@H{]}3C{[}N@@H+{]}1{[}C@@H{]}2{[}C@@H{]}3O\end{tabular} & $-2.42{ \cdot } 10^{1}$ \\
  & acetonitrile & \begin{tabular}[t]{@{}l@{}}{[}H{]}/N=C({\textbackslash}C){[}C@@H{]}1O{[}C@@H{]}({[}C-{]}(C\#N)C\\C\#C)OC{[}C@@H{]}1C{[}NH3+{]}\end{tabular} & $-5.61{ \cdot } 10^{1}$ \\
  & water & \begin{tabular}[t]{@{}l@{}}C{[}C@{]}12{[}NH2+{]}{[}C@H{]}1CNC1=N{[}C@@{]}3(CN)C\\N=C({[}O-{]}){[}C@H{]}3{[}C@@H{]}12\end{tabular} & $-7.06{ \cdot } 10^{1}$ \\
\cline{1-2} \multirow{5}{*}{EGP} & pentane & \begin{tabular}[t]{@{}l@{}}{[}C-{]}\#{[}N+{]}C(=O)C1=C=C=c2c3c4c(c5c(=O)\\c5c5sc6c7c(=O)c7c1c2c6c45)C(C(=O){[}N+\\{]}\#{[}C-{]})=C=C=3\end{tabular} & $-1.66{ \cdot } 10^{2}$ \\
  & acetonitrile & {[}NH3+{]}{[}C@@H{]}1CC{[}C@@H{]}(C(=O){[}O-{]})C1 & $-5.25{ \cdot } 10^{1}$ \\
  & water & {[}NH3+{]}{[}C@@H{]}1CC{[}C@@H{]}(C(=O){[}O-{]})C1 & $-6.71{ \cdot } 10^{1}$ \\
\midrule \multicolumn{4}{c}{max} \\
\midrule \multirow{3}{*}{\begin{tabular}[t]{@{}c@{}}GDB17\\amons\end{tabular}} & pentane & N=C=O.NC=S & $\phantom{-}9.69\phantom{{ \cdot } 10^{0}}$ \\
  & acetonitrile & NC/C=C{\textbackslash}C(=O)Br & $\phantom{-}5.35\phantom{{ \cdot } 10^{0}}$ \\
  & water & NC=O.S=C=NS & $\phantom{-}6.77\phantom{{ \cdot } 10^{0}}$ \\
\cline{1-2} \multirow{3}{*}{\begin{tabular}[t]{@{}c@{}}ZINC\\amons\end{tabular}} & pentane & C=C{[}C@H{]}(S)ON & $\phantom{-}1.04{ \cdot } 10^{1}$ \\
  & acetonitrile & O=CNC=O.{[}H{]}/N=N/{[}H{]} & $\phantom{-}9.39\phantom{{ \cdot } 10^{0}}$ \\
  & water & C=C(S)C{[}C@@H{]}1CS1 & $\phantom{-}8.47\phantom{{ \cdot } 10^{0}}$ \\
\cline{1-2} \multirow{4}{*}{GDB17} & pentane & CC/N=C1/NC=C2C1=NC(SC\#N)=C2O.N\#N & $\phantom{-}1.43{ \cdot } 10^{1}$ \\
  & acetonitrile & \begin{tabular}[t]{@{}l@{}}CC{[}C@@H{]}1{[}C@{]}2(OC)CC=CC(=O){[}C@{]}12N(C\\)C\end{tabular} & $\phantom{-}9.94\phantom{{ \cdot } 10^{0}}$ \\
  & water & Cc1c{[}nH{]}c(=O)sc(CCN)c(/N=C/N)cn1 & $\phantom{-}9.28\phantom{{ \cdot } 10^{0}}$ \\
\cline{1-2} \multirow{3}{*}{EGP} & pentane & CN(C){[}C@@{]}1({[}N+{]}(=O){[}O-{]})CSCCS1 & $\phantom{-}5.29\phantom{{ \cdot } 10^{0}}$ \\
  & acetonitrile & O=C(O)c1cccc(O)c1O & $\phantom{-}4.26\phantom{{ \cdot } 10^{0}}$ \\
  & water & CN(C){[}C@@H{]}1C=CSC(c2ccccc2)=C1 & $\phantom{-}9.17\phantom{{ \cdot } 10^{0}}$ \\
\bottomrule
\end{tabular}

\caption{Lowest ("min.") and highest ("max.") values of solvation energy $\Esolv$ in different subsets of SolQuest for pentane, acetonitrile, and water, along with the corresponding SMILES.}
\label{tab:SolQuest_extrema_solvation}
\end{table*}

\begin{table*}
 \centering
\begin{tabular}{lll}
\toprule
 \multirow{1}{*}{source} & \multirow{2}{*}{SMILES} & \multirow{2}{*}{$E_{\mathrm{tot}}$, a.u.} \\
 \multirow{1}{*}{dataset} &  &  \\
\midrule \multicolumn{3}{c}{min} \\
\midrule GDB17 amons & Br/C=N/C(Br)=C{\textbackslash}Br & $-7.89{ \cdot } 10^{3}$ \\
 ZINC amons & BrC(Br){[}SiH2{]}C(Br)Br & $-1.07{ \cdot } 10^{4}$ \\
 GDB17 & \begin{tabular}[t]{@{}l@{}}C{[}C@@H{]}(C)N(C)C(=O)c1c(Br)nc(Br)c(Br\\)c1Br\end{tabular} & $-1.09{ \cdot } 10^{4}$ \\
 EGP & \begin{tabular}[t]{@{}l@{}}BrC(Br)(Br)C1=CS/C(=C2/SC=C(C(Br)(Br\\)Br)S2)S1\end{tabular} & $-1.73{ \cdot } 10^{4}$ \\
\midrule \multicolumn{3}{c}{max} \\
\midrule GDB17 amons & C & $-4.05{ \cdot } 10^{1}$ \\
 ZINC amons & BC & $-6.60{ \cdot } 10^{1}$ \\
 GDB17 & CC1=CC(C)(C)COC1 & $-3.89{ \cdot } 10^{2}$ \\
 EGP & N & $-5.66{ \cdot } 10^{1}$ \\
\bottomrule
\end{tabular}

\caption{Lowest ("min.") and highest ("max.") values of total energy $\Etot$ (see explanation of \jsonkeyword{ENERGY} keyword in Subsec.~4.2 of the main text for definition) in different subsets of  SolQuest along with the corresponding SMILES.}
\label{tab:SolQuest_extrema_total_energy}
\end{table*}